\documentclass[preprint,review,3p,times,onecolumn,authoryear]{elsarticle}

\usepackage{amssymb}
\usepackage{multicol}

\usepackage{lineno}

\journal{Computers \& Geosciences}

\usepackage{xspace,color,amsmath}
\usepackage{amssymb}
\usepackage{enumitem}
\usepackage{multirow}
\setitemize{noitemsep,topsep=0pt,parsep=0pt,partopsep=0pt}

\newcommand{\SimPEG}{\textsc{SimPEG}\xspace}
\newcommand{\simpegEM}{\textsc{simpegEM}\xspace}

\newcommand{\Mesh}{\texttt{Mesh}\xspace}
\newcommand{\Survey}{\texttt{Survey}\xspace}

\newcommand{\Problem}{\texttt{Problem}\xspace}

\newcommand{\Regularization}{\texttt{Regularization}\xspace}
\newcommand{\DataMisfit}{\texttt{DataMisfit}\xspace}
\newcommand{\Optimization}{\texttt{Optimization}\xspace}
\newcommand{\InvProblem}{\texttt{InvProblem}\xspace}
\newcommand{\Inversion}{\texttt{Inversion}\xspace}

\newcommand{\Mapping}{\texttt{Mapping}\xspace}
\newcommand{\ExpMap}{\texttt{ExpMap}\xspace}
\newcommand{\PropMap}{\texttt{PropMap}\xspace}
\newcommand{\Sources}{\texttt{Sources}\xspace}
\newcommand{\Receivers}{\texttt{Receivers}\xspace}
\newcommand{\Fields}{\texttt{Fields}\xspace}

\newcommand{\TDEM}{\texttt{TDEM}\xspace}
\newcommand{\FDEM}{\texttt{FDEM}\xspace}
\newcommand{\NSEM}{\texttt{NSEM}\xspace}
\newcommand{\BaseEM}{\texttt{BaseEM}\xspace}

\newcommand{\InjectActiveCells}{\texttt{InjectActiveCells}\xspace}

\newcommand{\dpred}{\mathbf{d}_\text{pred}}

\newcommand{\sm}{\mathbf{s}_m}
\newcommand{\se}{\mathbf{s}_e}

\newcommand{\curl}{{\vec \nabla}\times\,}

\newcommand{\dcurl}{{\mathbf C}}

\newcommand{\M}{{\mathbf M}}
\newcommand{\MfMui}{{\M^f_{\boldsymbol{\mu^{-1}}}}}
\newcommand{\MfRho}{{\M^f_{\boldsymbol{\rho}}}}
\newcommand{\MeSigma}{{\M^e_{\boldsymbol{\sigma}}}}
\newcommand{\MeMu}{{\M^e_{\boldsymbol{\mu}}}}
\newcommand{\mui}{\mu^{-1}}

\usepackage{etoolbox}
\newtoggle{finaldraft}
\togglefalse{finaldraft}

\begin{document}

\begin{frontmatter}



{
\title{A framework for simulation and inversion in electromagnetics}

\author[corresponding,UBC]{Lindsey J. Heagy}
\author[UBC]{Rowan Cockett}
\author[UBC]{Seogi Kang}
\author[UBC]{Gudni K. Rosenkjaer}
\author[UBC]{Douglas W. Oldenburg}

\address[corresponding]{604-836-2715, lheagy@eos.ubc.ca}
\address[UBC]{Geophysical Inversion Facility, University of British Columbia}


\begin{abstract}

Simulations and inversions of electromagnetic geophysical data are paramount
for discerning meaningful information about the subsurface from these data.
Depending on the nature of the source electromagnetic experiments may be
classified as time-domain or frequency-domain. Multiple heterogeneous and
sometimes anisotropic physical properties, including electrical conductivity
and magnetic permeability, may need be considered in a simulation. Depending
on what one wants to accomplish in an inversion, the parameters which one
inverts for may be a voxel-based description of the earth or some parametric
representation that must be mapped onto a simulation mesh. Each of these
permutations of the electromagnetic problem has implications in a numerical
implementation of the forward simulation as well as in the computation of the
sensitivities, which are required when considering gradient-based inversions.
This paper proposes a framework for organizing and implementing
electromagnetic simulations and gradient-based inversions in a modular,
extensible fashion. We take an object-oriented approach for defining and
organizing each of the necessary elements in an electromagnetic simulation,
including: the physical properties, sources, formulation of the discrete
problem to be solved, the resulting fields and fluxes, and receivers used to
sample to the electromagnetic responses.  A corresponding implementation is
provided as part of the open source simulation and parameter estimation
project \SimPEG (http://simpeg.xyz). The application of the framework is
demonstrated through two synthetic examples and one field example. The first
example shows the application of the common framework for 1D time domain and
frequency domain inversions. The second is a field example that demonstrates a
1D inversion of electromagnetic data collected over the Bookpurnong Irrigation
District in Australia. The final example is a 3D example which shows how the modular implementation
is used to compute the sensitivity for a parametric model where a transmitter
is positioned inside a steel cased well.

\end{abstract}

\begin{keyword}
Geophysics, Numerical Modelling, Finite Volume, Sensitivities, Object Oriented
\end{keyword}
}

\end{frontmatter}



\section{Introduction}
\label{sec:intro}

The field of electromagnetic (EM) geophysics encompasses a diverse suite of
problems with applications across mineral and resource exploration,
environmental studies and geotechnical engineering. EM problems can be
formulated in the time or frequency domain. Sources can be grounded electric
sources or inductive loops driven by time-harmonic or transient currents, or
natural, plane wave sources, as in the case of the magnetotelluric method. The
physical properties of relevance include electrical conductivity, magnetic
permeability, and electric permittivity. These may be  isotropic, anisotropic,
and also frequency dependent. Working with electromagnetic data to discern
information about subsurface physical properties requires that we have
numerical tools for carrying out forward simulations and inversions that are
capable of handling each of these permutations.

The goal of the forward simulation is to solve a specific set of Maxwell's
equations and obtain a prediction of the EM responses. Numerical simulations
using a staggered grid discretization \citep{Yee1966}, have been extensively
studied in their application for finite difference, finite volume and finite
element approaches (c.f. \cite{newman1999, Haber2014a}), with many such
implementations being optimized for efficient computations for the context in
which they are being applied \citep{Haber2001, Key2007,  Kelbert2014,
Yang2014}.

Finding a model of the earth that is consistent with the observed data and
prior geologic knowledge is the `inverse problem'. It presupposes
that we have a means of solving the forward problem. The inverse problem is
generally solved by minimizing an  objective function that consists of a data
misfit and regularization, with a trade-off parameter controlling their
relative contributions. \citep{Tikhonov1977, Parker1980, Constable1987}.
Deterministic, gradient-based approaches to the inverse problem are
commonplace in EM inversions.   Relevance of the recovered inversion model is
increased by incorporating \emph{a priori} geologic information and
assumptions. This can be accomplished through, the regularization term
\citep{OldenburgTutorial, Constable1987} or parameterizing the inversion model
\citep{Pidlisecky2011, McMillan2015a, Kang2015}. Multiple data sets may be
considered through cooperative or joint inversions \citep{Haber1998,
McMillan2015}.

Each of these advances relies on a workflow and associated software
implementation. Unfortunately, each software implementation is typically developed as a stand-alone
solution. As a result, these advances are not readily interoperable with
regard to concepts, terminology, notation \emph{and} software.

The advancement of EM geophysical techniques and the expansion of their
application requires a flexible set of concepts and tools that are organized
in a framework so that researchers can more readily experiment with, and
explore, new ideas. For example, if we consider research questions within the
growing application of EM for reservoir characterization and monitoring in
settings with steel cased wells (cf. \cite{Hoversten2015, Um2015, Commer2015,
cuevas2014, Hoversten2014, Pardo2013}), the numerical tools employed must
enable investigation into factors such as the impact of variable magnetic
permeability \citep{wuhabashy1994, Heagy2015} and casing integrity
\citep{brill2012} on electromagnetic signals. Various modelling approaches in
both time and frequency domain simulations are being explored, these include
employing highly-refined meshes \citep{Commer2015}, using cylindrical symmetry
\citep{Heagy2015} or approximating the casing on a coarse-scale
\citep{Um2015}, possibly 3D anisotropic approximations
\citep{CaudilloMata2014}. Beyond forward simulations that predict EM
responses, to enable the interpretation of field data with these tools
requires that machinery to address the inverse problem and experiment with
approaches for constrained and/or time lapse inversions be in place
\citep{Devriese2016, Marsala2015}. Typically, addressing each of these
complexities would require a custom implementation, particularly for the
frequency domain and time domain simulations, although aspects, such as
physical properties, are common to both. Inconsistencies between
implementations and the need to implement a custom solution for each type of
EM method under consideration presents a significant barrier to a researcher's
ability to experiment with and extend ideas.

Building from the body of work on EM geophysical simulations and inversions,
the aim of our efforts is to identify a common, modular framework suitable
across the suite of electromagnetic problems. This conceptual organization has
been tested and developed through a numerical implementation. The
implementation is modular in design with the expressed goal of affording
researchers the ability to rapidly adjust, interchange, and extend elements.
By developing the software in the open, we also aim to promote an open dialog
on approaches for solving forward and inverse problems in EM geophysics.

The implementation we describe for EM forward and inverse problems extends a
general framework for geophysical simulation and gradient based inverse
problems, called \SimPEG \citep{Cockett2015}. The implementation of \SimPEG is
open-source, written in Python and has dependencies on the standard numerical
computing packages NumPy, SciPy, and Matplotlib \citep{numpy, scipy,
matplotlib}. The contribution described in this paper is the implementation of
the physics engine for problems in electromagnetics, including the forward
simulation and calculation of the sensitivities (\simpegEM). Building within the \SimPEG
ecosystem has expedited the development process and allowed developments to be
made in tandem with other applications (http://simpeg.xyz). \simpegEM aspires
to follow best practices in terms of documentation, testing, continuous
integration using the publically available services Sphinx, Travis CI, and
Coveralls \citep{Sphinx, Travis, Coveralls}. As of the writing of this paper,
when any line of code is changed in the open source repository, over 3 hours
of testing is completed; documentation and examples are also tested and
automatically updated (http://docs.simpeg.xyz). We hope these practices
encourage the growth of a community and collaborative, reproducible software
development in the field of EM geophysics.

\bigskip

The paper is organized as follows. To provide context for the structure and
implementation of \simpegEM, we begin with a brief overview of the \SimPEG
inversion framework as well as the governing equations for electromagnetics in
Section~\ref{sec:Background}. In Section~\ref{sec:Motivation}, we discuss the
motivating factors for the EM framework, and in
Section~\ref{sec:Implementation}, we discuss the framework and implementation
of the forward simulation and calculation of sensitivities in \simpegEM. We
demonstrate the implementation with two synthetic examples and one field
example in Section~\ref{sec:Examples}. The first example shows the
similarities between the time and frequency domain implementations for a 1D
inversion. In the second example, we invert field data from the Bookpurnong
Irrigation district in Australia. The final example is a 3D example that demonstrates how the
modular implementation is used to compute the sensitivity for a parametric
model of a block in a layered space where a transmitter is positioned inside a
steel cased well.


\section{Background}
\label{sec:Background}

We are focused on geophysical inverse problems in electromagnetics (EM), that
is, given EM data, we want to find a model of the earth that explains those
data and satisfies prior assumptions about the geologic setting. We follow the
\SimPEG framework, shown in Figure \ref{fig:SimPEG}, which takes a gradient-based
approach to the inverse problem \citep{Cockett2015}. Inputs to the
inversion are the data and associated uncertainties, a description of the
governing equations, as well as prior knowledge and assumptions about the
model. With these defined, the \SimPEG framework accomplishes two main
objectives:
\begin{enumerate}
\item the ability to forward simulate data and compute sensitivities (Forward Simulation - outlined in green in Figure~\ref{fig:SimPEG}),
\item the ability to assess and update the model in an inversion (Inversion Elements and Inversion as Optimization - outlined in red in Figure~\ref{fig:SimPEG}).
\end{enumerate}

{
\begin{figure}[htb!]
    \centering
    \includegraphics[width=7.5cm]{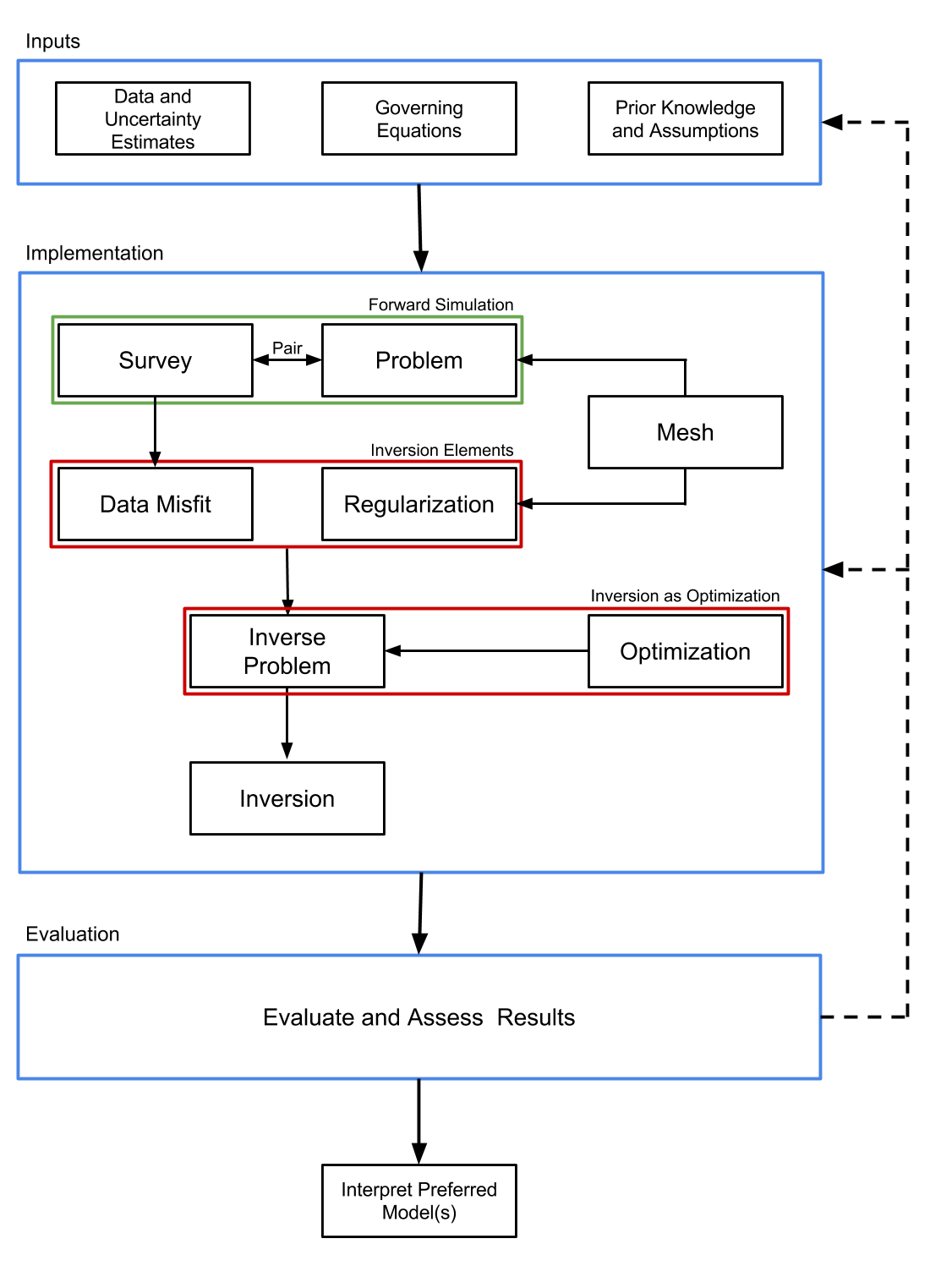}
\caption{Inversion approach using the \SimPEG framework. Adapted from \cite{Cockett2015}}
\label{fig:SimPEG}
\end{figure}
}

The implementation of the framework is organized into the self-contained
modules shown in Figure~\ref{fig:SimPEG}; each module is defined as a base-class
within \SimPEG. The \Mesh provides the discretization and numerical
operators. These are leveraged by the \Problem, which is the numerical physics
engine; the \Problem computes fields and fluxes when provided a model and \Sources. The
\Sources are specified in the \Survey, as are the \Receivers. The \Receivers
take the \Fields computed by the \Problem and evaluate them at the receiver
locations to create predicted data. Each action taken to compute data, when
provided a model, has an associated derivative with respect to the model;
these components are assembled to create the sensitivity. Having the ability
to compute both predicted data and sensitivities accomplishes the first
objective.

To accomplish the second objective of assessing and updating the model in the context of the
data and our assumptions, we consider a gradient-based approach to the
inversion. For this, we specify an objective function which generally consists of a
\DataMisfit and \Regularization. The \DataMisfit is a metric that evaluates
the agreement between the observed and predicted data, while the
\Regularization is a metric constructed to assess the model's agreement with
assumptions and prior knowledge. These are combined with a trade-off parameter
to form a mathematical statement of the \InvProblem, an optimization problem.
The machinery to update the model is provided by the \Optimization. An
\Inversion brings all of the elements together and dispatches \texttt{Directives}
for solving the \InvProblem. These \texttt{Directives} are
instructions that capture the heuristics for solving the inverse problem; for
example, specifying a target misfit that, once reached, terminates the
inversion, or using a beta-cooling schedule that updates the value of the
trade-off parameter between the \DataMisfit and \Regularization (cf.
\cite{parker1994geophysical, OldenburgTutorial} and references within).

The output of this process is a model that must be assessed and evaluated
prior to interpretation; the entire process requires iteration by a human,
where underlying assumptions and parameter choices are re-evaluated and
challenged. Be it in resource exploration, characterization or development;
environmental remediation or monitoring; or geotechnical applications -- the
goal of this model is to aid and inform a complex decision.

\bigskip

Here we note that the inversion framework described above is agnostic to the
type of forward simulation employed, provided the machinery to solve the
forward simulation and compute sensitivities is implemented. Specific to the
EM problem, we require this machinery for Maxwell's equations. As such, we
focus our attention  on the \texttt{Forward Simulation} portion of the
implementation for the EM problem and refer the reader to
\cite{Cockett2015} and \cite{OldenburgTutorial} for a more complete discussion
of  inversions.


\subsection{Governing Equations}
\label{sec:GoverningEquations}

Maxwell's equations are the governing equations of electromagnetic problems.
They are a set of coupled partial differential equations that connect electric
and magnetic fields and fluxes. We consider the quasi-static regime, ignoring
the contribution of displacement current \citep{Ward1988, telford1990applied,
Haber2014a} \footnote{In most geophysical electromagnetic surveys, low frequencies or late-time measurements are employed. In these scenarios $\sigma \gg \varepsilon_0 \omega$ (eg. conductivities are typically less than 1S/m, $\varepsilon_0 = 8.85 \times 10^{-12} F/m$ and frequencies considered are generally less than $10^5$ Hz), so displacement current can safely be ignored.}

We begin by considering the first order quasi-static EM problem in time,
\begin{equation}
\begin{split}
\curl \vec{e} + \frac{\partial \vec{b}}{\partial t} = \vec{s}_m \\
\curl \vec{h} - \vec{j} = \vec{s}_e
\end{split}
\label{eq:MaxwellBasicTime}
\end{equation}
where $\vec{e}$, $\vec{h}$ are the electric and magnetic fields, $\vec{b}$ is the magnetic flux density, $\vec{j}$ is the current density, and $\vec{s}_m$, $\vec{s}_e$ are the magnetic and electric source terms. $\vec{s}_e$ is a physical, electric current density, while $\vec{s}_m$ is ``magnetic current density''. Although $\vec{s}_m$ is unphysical, as continuity of the magnetic current density would require magnetic monopoles, the definition of a magnetic source term can be a useful construct, as we will later demonstrate in Section~\ref{sec:Implementation} (see also \cite{Ward1988}).

By applying the Fourier Transform (using the $e^{i\omega t}$ convention), we can write Maxwell's equations in the frequency domain:
\begin{equation}
\begin{split}
\curl \vec{E} + i\omega\vec{B} = \vec{S}_m \\
\curl \vec{H} - \vec{J} = \vec{S}_e
\end{split}
\label{eq:MaxwellBasicFreq}
\end{equation}
where we use capital letters to denote frequency domain variables. The fields and fluxes are  related through the physical properties: electrical conductivity $\sigma$, and magnetic permeability $\mu$, as described by the constitutive relations
\begin{equation}
\begin{split}
\vec{J} = \sigma \vec{E} \\
\vec{B} = \mu \vec{H}
\end{split}
\label{eq:ConstitutiveRelations}
\end{equation}
The physical properties, $\sigma$ and $\mu$ are generally distributed and heterogeneous. For isotropic materials, $\sigma$ and $\mu$ are scalars, while for anisotropic materials they are $3\times3$ symmetric positive definite tensors. The same constitutive relations can be applied in the time domain provided that the physical properties, $\sigma$, $\mu$ are not frequency-dependent.

In an EM geophysical survey, the sources provide the input energy to excite responses that depend on the physical property distribution in the earth. These responses, electric and magnetic fields and fluxes, are sampled by receivers to give the observed data. The simulation of Maxwell's equations may be conducted in either the time or frequency domain, depending on the nature of the source; harmonic waveforms are naturally  represented in the frequency domain, while transient waveforms are better described in the time domain.

The aim of the inverse problem is to find a model, $\mathbf{m}$ (which may be a voxel-based or a parametric representation) that is consistent with observed data and with prior knowledge and assumptions about the model. Addressing the inverse problem using a gradient-based approach requires two abilities of the forward simulation: (1) the ability to compute predicted data given a model
\begin{equation}
\dpred = \mathcal{F}[\mathbf{m}]
\label{eq:dpred}
\end{equation}
and (2) the ability to compute or access the sensitivity, given by
\begin{equation}
    \mathbf{J}[\mathbf{m}] = \frac{d \mathcal{F}[\mathbf{m}]}{ d \mathbf{m}}.
    \label{eq:sensitivity}
\end{equation}
To employ second order optimization techniques, we also require  the adjoint of the sensitivity, $\mathbf{J}^\top$. These two elements, when combined into the \SimPEG framework, enable data to be simulated and gradient-based inversions to be run. As such, this work benefits from other peoples' contributions to the underlying inversion machinery, including: discrete operators on a variety of meshes, model parameterizations, regularizations, optimizations, and inversion directives \citep{Cockett2015}.


\section{Motivation}
\label{sec:Motivation}

The motivation for the development of this framework is that it be a resource
for researchers in the field of electromagnetic geophysics. To best serve this
goal, we require a framework that is modular and extensible in order to enable
exploration of ideas. An associated numerical implementation is essential for
this work to be tested and acted upon. As such, we provide a tested,
documented, fully open-source software implementation of the framework (under the permissive MIT license).

Specific to the EM problem, we require the implementation of Maxwell's
equations in both the time domain and frequency domain. The implementation
must allow for variable electrical conductivity and magnetic permeability,
anisotropic physical properties; various model parameterizations of the
physical properties (e.g. voxel  log-conductivity or parametric
representations); a range of sources including wires, dipoles, natural
sources; variable receiver types; variable formulations of Maxwell's
equations; solution approaches such as using a primary-secondary formulation; and the flexibility
to work with and move between a variety of meshes such as tensor,
cylindrically symmetric, curvilinear, and octree discretizations. Furthermore,
the sensitivity computation must be flexible enough to be computed for any
sensible combination of these approaches. In the following section, we will
outline the framework we have used to organize and implement these ideas.


\section{Simulation Framework}
\label{sec:Implementation}

The aim of the forward simulation is to compute predicted data, $\dpred$,
when provided with an inversion model\footnote{We use the term \emph{inversion model}
to describe a parameterized representation of the earth (e.g. voxel-based or
parametric), even if the model is solely used for forward modelling, its form
sets the context for the inverse problem and the parameter-space that is to be
explored.}, $\mathbf{m}$ and \Sources. \simpegEM contains implementations for
both time domain (\TDEM) and frequency domain (\FDEM) simulations, allowing
data from commonly used EM methods to be simulated.

The framework we follow to perform the forward simulation is shown in
Figure~\ref{fig:simpegFwd}; it consists of two overarching categories:
\begin{enumerate}
\item the \Problem, which is the implementation of the governing equations,
\item the \Survey, which provides the source(s) to excite the system as well as the receivers to samples the fields and produce predicted data at receiver locations.
\end{enumerate}
Here, we provide a brief overview of each of the components, and discuss them
in more detail in the sections that follow.
{
\begin{figure}[htb!]
    \centering
    \includegraphics[width=0.8\textwidth]{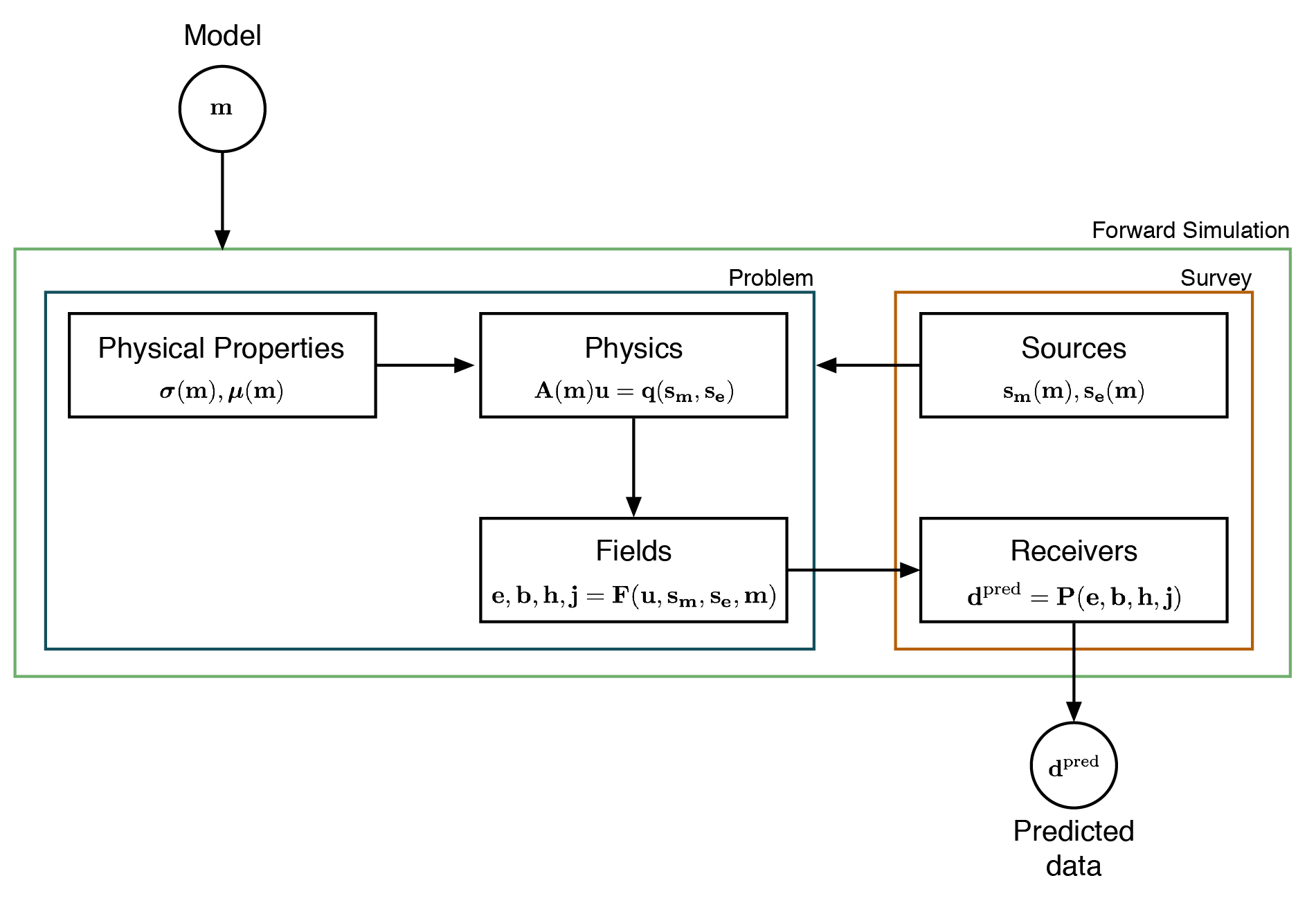}
\caption{Forward simulation framework.}
\label{fig:simpegFwd}
\end{figure}
}

\begin{figure}[htb!]
    \centering
    \includegraphics[width=0.55\textwidth]{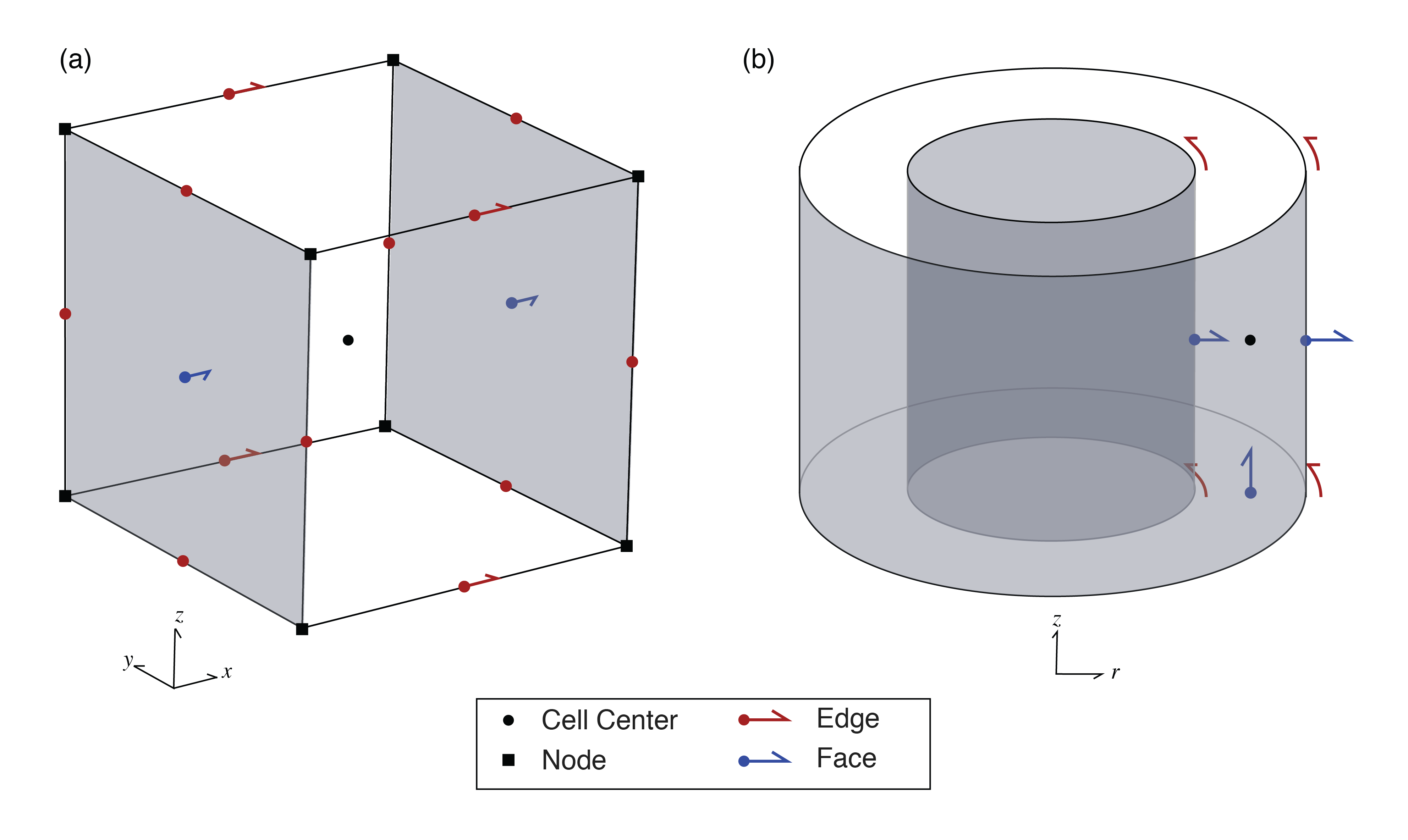}
    \caption{Location of variables in the finite volume implementation for both a unit cell in (a) cartesian and (b) cylindrical coordinates (after \cite{Heagy2015})}
\label{fig:finiteVolumes}
\end{figure}

The `engine' of the forward simulation is the physics; it contains the
machinery to solve the system of equations for EM fields and fluxes in the
simulation domain when provided with a description of the physical properties
and sources. In general, the physics engine may be an analytic or numeric
implementation of Maxwell's equations. Here, we focus our attention on the
numerical implementation using a standard staggered-grid finite volume
approach, requiring that the physical properties, fields, fluxes and sources
be defined on a mesh (cf. \cite{Haber2014a, Hyman2002, Hyman1999, Yee1966}).
We discretize fields on edges, fluxes on faces and physical properties in cell
centers, as shown in Figure \ref{fig:finiteVolumes}. To construct the
necessary differential and averaging operators, we leverage the \Mesh class
within \SimPEG \citep{Cockett2015, FV_Tutorial}.

To compute electromagnetic responses, the forward simulation requires the
definition of a physical property model describing the electrical conductivity
($\boldsymbol{\sigma}$) and magnetic permeability ($\boldsymbol{\mu}$) on the
simulation mesh, as well as discrete representations of the sources used to
excite EM responses ($\boldsymbol{s_e}, \boldsymbol{s_m})$. Often in solving
an inverse problem, the model which one inverts for (the vector $\mathbf{m}$),
is some discrete representation of the earth that is decoupled from the
physical property model. This decoupling requires the definition of a
\texttt{Mapping} capable of translating $\mathbf{m}$ to physical properties on
the simulation mesh. For instance, if the inversion model is chosen to be
log-conductivity, an exponential mapping is required to obtain electrical
conductivity (i.e. $\boldsymbol{\sigma} = \mathcal{M}(\mathbf{m}$)). To
support this abstraction, \SimPEG provides a number of extensible \Mapping
classes \citep{Cockett2015, Kang2015}.

With both the physical property model and the source specified, we define and
solve the physics, a Maxwell system of the form
\begin{equation}
\mathbf{A}(\mathbf{m})\mathbf{u} = \mathbf{q}(\sm,\se),
    \label{eq:physics}
\end{equation}
for an electric or magnetic field or flux. Here, $\mathbf{A}$ is the system
matrix that may eliminate a field or flux to obtain a second-order system in a single field
or flux, $\mathbf{u}$, the solution vector.  Correspondingly, the vector
$\mathbf{q}$ is the second order right-hand-side. Note, if there are necessary
manipulations to make equation \ref{eq:physics} easier to solve numerically
(e.g. symmetry) we can add these here; doing so has no effect on the
derivative. The remaining fields and fluxes can be computed from $\mathbf{u}$
anywhere in the simulation domain, through an operation of the form
\begin{equation}
    \mathbf{f} = \mathbf{F}(\mathbf{u(m)}, \mathbf{s}_e(\mathbf{m}), \mathbf{s}_m(\mathbf{m}), \mathbf{m})
\label{eq:fields}
\end{equation}
where $\mathbf{f}$ is conceptually a vector of \emph{all} of the fields and
fluxes (i.e. $\mathbf{e}$, $\mathbf{b}$, $\mathbf{h}$ and $\mathbf{j}$). This
vector is never stored in the implementation, instead the fields are computed
on demand through the subset of stored solution vectors ($\mathbf{u}$). From the
computed fields ($\mathbf{f}$), predicted data are created by the \Receivers
through an operation of the form
\begin{equation}
    \dpred = \mathbf{P}(\mathbf{f})
\label{eq:data}
\end{equation}
In the simplest case, the action of $\mathbf{P}$ selects the component of interest
and interpolates the fields to the receiver locations, more involved cases
could include the computation of ratios of fields, as is the case for
impedance or tipper data. Obtaining predicted data from the framework
concludes the forward simulation.

The same framework is employed for both time domain (\TDEM) and frequency
domain (\FDEM) implementations within \simpegEM. In the case of the \FDEM
implementation, the matrix $\mathbf{A}(\mathbf{m})$ and the solution vector
$\mathbf{u}$ represent all frequencies. As these frequencies are independent
(i.e. a block diagonal matrix,
\includegraphics[width=0.02\textwidth]{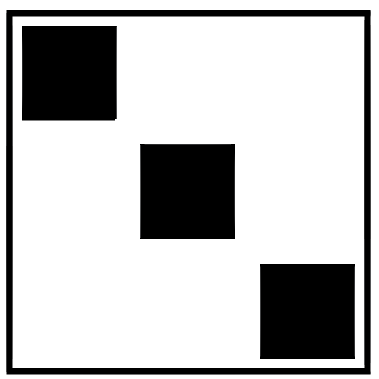}),  each frequency can
be solved independently. In the \TDEM code, the matrix
$\mathbf{A}(\mathbf{m})$ and the solution vector $\mathbf{u}$ represent all
timesteps \citep{Oldenburg2013, Haber2014a} and take the form of a  lower
triangular block matrix (bidiagonal in the case of Backward Euler,
\includegraphics[width=0.02\textwidth]{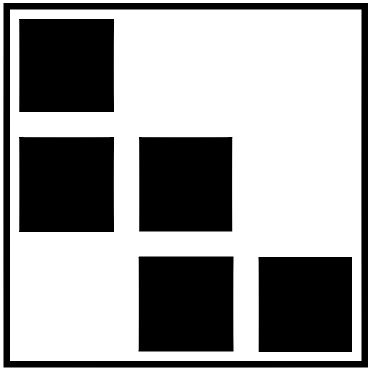}),  meaning the
computation of each time-step depends on previous time-steps. The form of
these matrices will be discussed further in the Physics section (Section \ref{sec:Physics})

To perform a gradient-based inversion, we require the sensitivity of the data
with respect to the inversion model, thus, each action taken to calculate data
from the model must have an associated derivative. The full sensitivity is a
dense matrix and is expensive to form and store, but when the optimization
problem is solved using an iterative optimization approach, it does not need
to be explicitly formed; all that is required are products and adjoint-products
with a vector. We treat this using a modular approach so that individual
elements of the framework can be rapidly interchanged or extended. The process
we follow to compute matrix-vector products with the sensitivity is shown with red arrows in
Figure~\ref{fig:Jvec} (b). The sensitivity-vector product $\mathbf{Jv}$ is built in stages by taking matrix
vector products with the relevant derivatives in each module, starting with the derivative of the physical property with respect to the model.
The product with the adjoint is similarly shown in Figure~\ref{fig:Jvec} (c) starting with the adjoint of the receiver operation.

{
\begin{figure}[htb!]
    \centering
    \includegraphics[width=\textwidth]{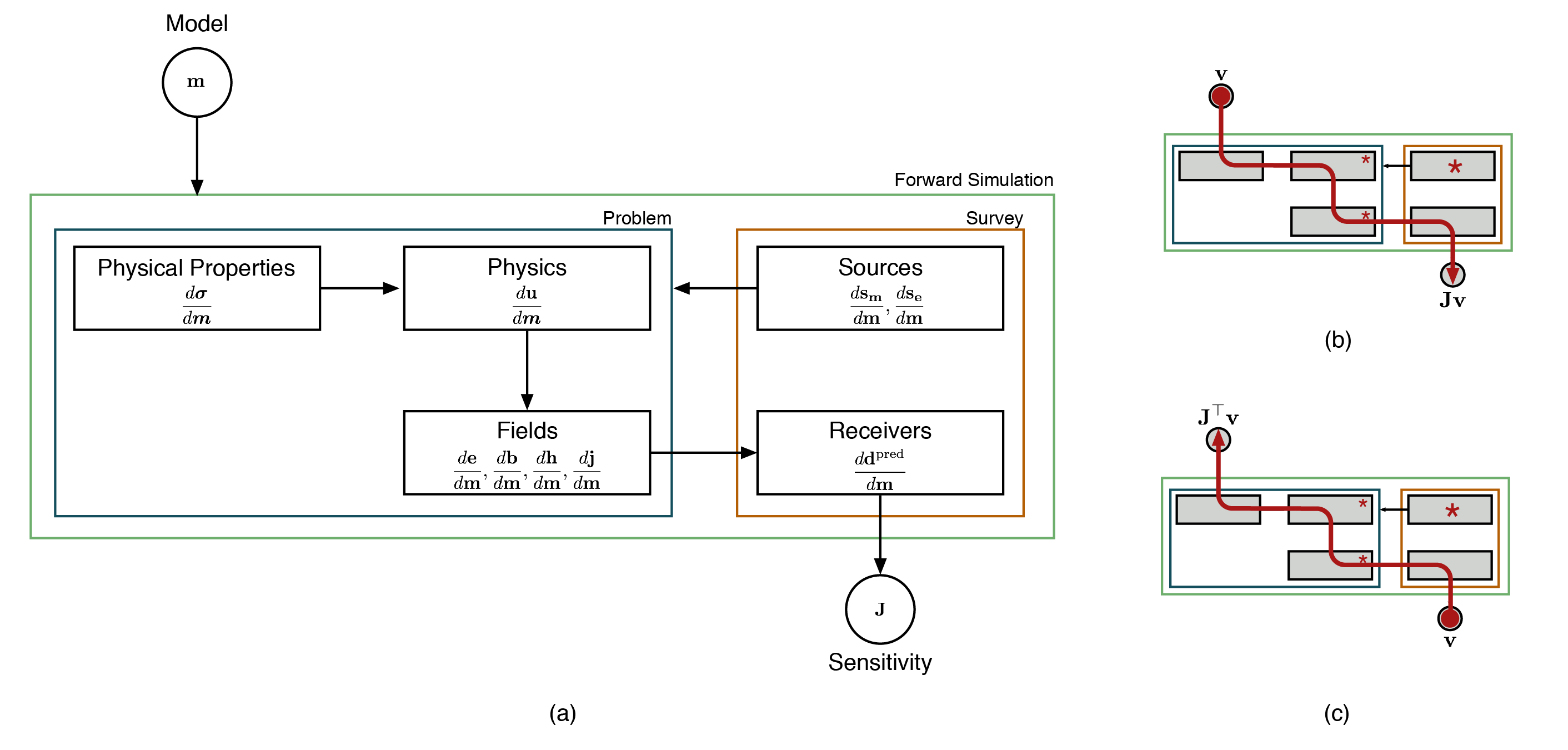}
    \caption{(a) Contributions of each module to the sensitivity. (b) process for computing $\mathbf{J} \mathbf{v}$ and (c) $\mathbf{J}^{\top}\mathbf{v}$; stars indicate where the source derivatives are incorporated.}
\label{fig:Jvec}
\end{figure}
}

Using electrical conductivity, $\boldsymbol{\sigma}$, as the only active
property described by the inversion model $\mathbf{m}$ for brevity, the
sensitivity takes the form
\begin{equation}
\mathbf{J}[\mathbf{m}] =
    \frac{\text{d}\mathbf{P}(\mathbf{f})}{\text{d}\mathbf{f}}
    \frac{\text{d}\mathbf{f}}{\text{d}\boldsymbol{\sigma}}
    \frac{\text{d}\boldsymbol{\sigma}}{\text{d}\mathbf{m}}
=
    \underbrace{
    \frac{\text{d}\mathbf{P}(\mathbf{f})}{\text{d}\mathbf{f}}
    }_{\text{Receivers}}
    \underbrace{
    \left(\vphantom{\frac{\partial\mathbf{f}}{\partial\mathbf{u}}} \right.
\frac{\partial\mathbf{f}}{\partial\mathbf{u}} \overbrace{\frac{\text{d}\mathbf{u}}{\text{d}\boldsymbol{\sigma}}}^{\text{Physics}}
        + \frac{\partial\mathbf{f}}{\partial\sm} \overbrace{\frac{\text{d}\sm}{\text{d}\boldsymbol{\sigma}}}^{\text{Sources}}
        + \frac{\partial\mathbf{f}}{\partial\se} \overbrace{\frac{\text{d}\se}{\text{d}\boldsymbol{\sigma}}}^{\text{Sources}}
        + \frac{\partial\mathbf{f}}{\partial\boldsymbol{\sigma}}
    \left. \vphantom{\frac{\partial\mathbf{f}}{\partial\mathbf{u}}}\right)
     }_{\text{Fields}}
    \underbrace{\frac{\text{d}\boldsymbol{\sigma}}{\text{d}\mathbf{m}}}_{\text{Properties}}
    \label{eq:generalsensitivity}
\end{equation}
The annotations denote which of the elements shown in Figure~\ref{fig:Jvec}
are responsible for computing the respective contribution to the sensitivity.
If the model provided is in terms of $\boldsymbol{\mu}$ or a source/receiver
location, this property replaces the role of $\boldsymbol{\sigma}$. The
flexibility to invoke distinct properties of interest (e.g. $\sigma$, $\mu$,
source location, etc.) in the inversion requires quite a bit of `wiring' to
keep track of which model parameters are associated with which properties;
this is achieved through a property mapping or \PropMap (physical properties,
location properties, etc.) within \SimPEG.

Although typically the source terms do not have model dependence and thus
their derivatives are zero, the derivatives of $\se$ and $\sm$ must be
considered in a general implementation. For example, if one wishes to use a
primary-secondary approach, where source fields are constructed by solving a
simplified problem, the source terms may have dependence on the model meaning
their derivatives have a non-zero contribution to the sensitivity (c.f.
\cite{Coggon1971, Haber2014a, Heagy2015}); this will be
demonstrated in the Casing Example in Section \ref{sec:steelcasing}.

The derivative of the solution vector $\mathbf{u}$ with respect to the model
is found by implicitly taking the derivative of equation \ref{eq:physics} with
respect to $\mathbf{m}$, giving
\begin{equation}
    \frac{\text{d}\mathbf{u}}{\text{d}\mathbf{m}} = \mathbf{A}^{-1}(\mathbf{m})
    \left(\vphantom{\frac{\partial \mathbf{A}(\mathbf{m}) \mathbf{u^{\text{fix}}}}{\partial \mathbf{m}}
        } \right.
    - \underbrace{\frac{\partial \mathbf{A}(\mathbf{m}) \mathbf{u^{\text{fix}}}}{\partial \mathbf{m}}
        }_{\texttt{getADeriv}}
        +
\underbrace{
         \frac{\partial\mathbf{q}}{\partial\sm} \frac{\text{d} \sm}{\text{d} \mathbf{m}}
        + \frac{\partial\mathbf{q}}{\partial\se} \frac{\text{d} \se}{\text{d} \mathbf{m}}
        + \frac{\partial\mathbf{q}}{\partial\mathbf{m}}
        }_{\texttt{getRHSDeriv}}
    \left. \vphantom{\frac{\text{d}\mathbf{q}}{\text{d}\sm} \frac{\text{d} \sm}{\text{d} \mathbf{m}}
        + \frac{\text{d}\mathbf{q}}{\text{d}\se} \frac{\text{d} \se}{\text{d} \mathbf{m}}
        + \frac{\partial\mathbf{q}}{\partial\mathbf{m}}}
    \right)
\label{eq:dudm}
\end{equation}
The annotations below the equation indicate the methods of the \Problem class
that are responsible for calculating the respective derivatives. Typically the
model dependence of the system matrix is through the physical properties (i.e.
$\sigma$, $\mu$). Thus, to compute derivatives with respect to $\mathbf{m}$,
the derivatives are first taken with respect to $\boldsymbol{\sigma}$ and the dependence of
$\boldsymbol{\sigma}$ on $\mathbf{m}$ is treated using chain rule. The chain
rule dependence is computed and tested automatically in \SimPEG using the
composable \Mapping classes.

In the following sections, we discuss the implementation of
elements shown in Figure~\ref{fig:simpegFwd} and highlight their contribution
to the forward simulation and calculation of the sensitivity. We begin by
discussing the inversion model and its relationship to the physical properties
(Section \ref{sec:ModelPhysProps}), move on to the core of the forward
simulation, the Physics (Section \ref{sec:Physics}), and to how Sources which excite the
system are defined (Section \ref{sec:Sources}). Following these, we then discuss how Fields are
calculated everywhere in the domain (Section \ref{sec:Fields}) and how they are evaluated by the
Receivers to create predicted data (Section \ref{sec:Receivers}). We conclude this section with a
Summary and discussion on testing (Section \ref{sec:Summary}).


\subsection{Model and Physical Properties}
\label{sec:ModelPhysProps}

For all EM problems, we require an inversion model
that can be mapped to meaningful physical properties in the discretized
Maxwell system. Typically, we consider the model to be a description of the
electrical conductivity distribution in the earth. Often, the model is taken
to be log-conductivity, in which case, an exponential mapping is required
(\ExpMap) to convert the model to electrical conductivity. The inversion model
may be defined on a subset of a mesh and referred to as an  `active cell' model.
For instance, air cells may be excluded and only the subsurface considered; in
this case an \InjectActiveCells map is used to inject the active model into
the full simulation domain. In the case of a parametric inversion, the
inversion model is defined on a domain that is independent of the forward
modelling mesh and the mapping takes the parametric representation and defines
a physical property on the forward modelling mesh  (e.g. a gaussian ellipsoid
defined geometrically) \citep{MaokunLi2010, Pidlisecky2011, McMillan2015,
Kang2015}. Maps can be composed, for instance, a layered, 1D log conductivity
model defined only in the subsurface may be mapped to a 2D cylindrical
\texttt{Mesh}, as shown in Figure \ref{fig:mapping}.
{
\begin{figure}[htb!]
    \centering
    \includegraphics[width=\textwidth]{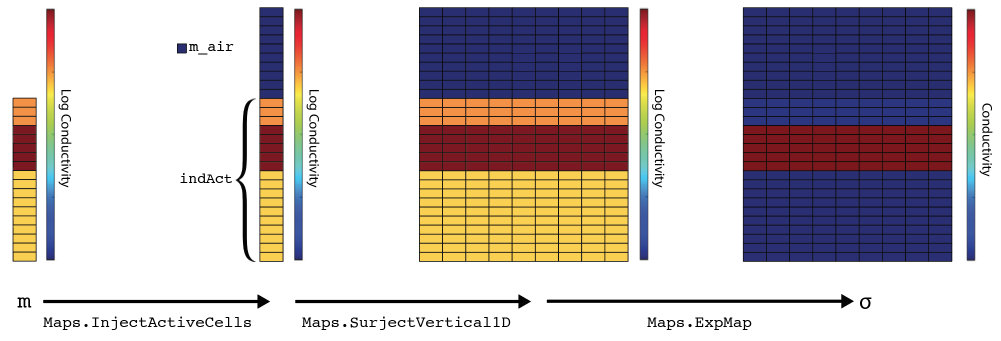}
    \caption{Mapping an inversion model, a 1D layered, log conductivity model defined below the surface, to electrical conductivity defined in the full simulation domain.}
\label{fig:mapping}
\end{figure}
}\noindent
{\scriptsize
\begin{verbatim}
    import numpy as np
    from SimPEG import Mesh, Maps
    mesh = Mesh.CylMesh([20, 20])    # SimPEG cylindrically symmetric mesh
    m_air = np.log(1e-8)             # value of the model in the air cells
    indAct = mesh.vectorCCz < 0.0    # define active cells to be subsurface only
    mapping = ( Maps.ExpMap(mesh) *
                Maps.SurjectVertical1D(mesh) *
                Maps.InjectActiveCells(mesh, indAct, m_air, nC=mesh.nCz) )
\end{verbatim}
}\noindent
In the code above, the `multiplication' performs the composition of the
mappings. For the contribution of this action to the sensitivity, the
derivative of the electrical conductivity with respect to the model is
computed using the chain rule for the composed maps (cf. \cite{Kang2015,
Heagy2014a}). During an inversion, the electrical conductivity on the
simulation mesh associated with the current inversion model and its derivative
are accessed through the \texttt{BaseEMProblem}, which is inherited by both
the \TDEM and \FDEM problems. In some cases, variable magnetic permeability
must be considered; this is accomplished through a property mapping
(\PropMap). The \PropMap handles the organization and independent mappings of
distinct physical properties (i.e. $\boldsymbol{\sigma}$, $\boldsymbol{\mu}$).


\subsection{Physics}
\label{sec:Physics}

To formulate a system of equations from Maxwell's equations in time
(equation~\ref{eq:MaxwellBasicTime}) or frequency
(equation~\ref{eq:MaxwellBasicFreq}) that can be solved numerically using a
finite volume approach, we require a statement of the problem in terms of two
equations with two unknowns, one of which is a field (discretized on edges),
and the other a flux (discretized on faces). Thus, we can consider either the
E-B formulation, or the H-J formulation. For the frequency-domain problem, we
can discretize the electric field, $\vec{e}$, on edges, the magnetic flux,
$\vec{b}$, on faces, physical properties $\sigma$ and $\mu^{-1}$ at cell
centers, and the source terms $\vec{s}_m$ and $\vec{s}_e$ on faces and edges,
respectively (see Figure~\ref{fig:finiteVolumes}). Doing so, we obtain the
discrete system:
\begin{equation}
    \begin{split}
        \dcurl \mathbf{e} + i\omega\mathbf{b} = \sm \\
        \dcurl^\top \MfMui \mathbf{b} - \MeSigma \mathbf{e} = \se
    \end{split}
    \label{eq:EB_FDEM}
\end{equation}
where $\dcurl$ is the discrete edge curl, $\MfMui$ is the face inner-product
matrix for $\boldsymbol{\mui}$, $\MeSigma$ is the edge inner-product matrix
for $\boldsymbol{\sigma}$; these inner product matrices can be computed for
isotropic, diagonally anisotropic or fully anisotropic physical properties
using operators within \SimPEG's \Mesh class \citep{Cockett2015, FV_Tutorial}.

Note that
the source-term $\se$ is an integrated quantity. Alternatively, the H-J
formulation discretizes $\vec{h}$ on edges, $\vec{j}$ on faces, $\rho$ and
$\mu$ at cell centers, and the source terms $\vec{s}_m$, $\vec{s}_e$ on edges
and faces, respectively, giving
\begin{equation}
    \begin{split}
        \dcurl^\top \MfRho \mathbf{j} + i\omega\MeMu\mathbf{h} = \sm \\
        \dcurl \mathbf{h} - \mathbf{j} = \se.
    \end{split}
    \label{eq:HJ_FDEM}
\end{equation}
Similarly, $\sm$ is an integrated quantity. In a full 3D simulation, the
electric and magnetic contributions for the two formulations are merely
staggered from one another. However, if using an assumption of cylindrically
symmetry, the appropriate formulation must be used to simulate either
rotational electric or magnetic contributions \citep{Heagy2015}. For both the
basic \FDEM and \TDEM implementations, natural boundary conditions ($\mathbf{b}
\times \hat{\mathbf{n}} = 0 ~ \forall \vec{x} \in \partial \Omega$ in E-B
formulation or $\mathbf{j} \times \hat{\mathbf{n}} = 0 ~ \forall \vec{x}
\in \partial \Omega$ in H-J formulation), in which the fields are assumed to
have decayed to a negligible value at the boundary, are employed to construct
the differential operators, the framework and implementation are however,
extensible to consider other boundary conditions (cf. \cite{Haber2014a,
Rivera-rios2014}).

In order to solve either equation \ref{eq:EB_FDEM} or equation \ref{eq:HJ_FDEM}, we eliminate
one variable and solve the second order system. This elimination is performed
by the FDEM problem classes. For instance, in \FDEM \texttt{Problem\_e}, we
eliminate $\mathbf{b}$ and obtain a second order system in $\mathbf{e}$
\begin{equation}
\underbrace{
\overbrace{
\left(\dcurl^\top \MfMui \dcurl + i\omega\MeSigma\right)
}^{\mathbf{A}(m)}}_{\texttt{getA}}
\overbrace{\vphantom{\left(\MfMui\right)}
            \mathbf{e}}^{\mathbf{u}}
=
\underbrace{ \vphantom{\left(\MfMui\right)}
\overbrace{
\dcurl^\top \MfMui \sm - i \omega \se
}^{\mathbf{q(s_m, s_e)}}
}_{\texttt{getRHS}}
\label{eq:problem_e}
\end{equation}

\FDEM \texttt{Problem\_e} has methods \texttt{getA} and \texttt{getRHS} to construct the system
{\scriptsize
\begin{multicols}{2}
\begin{verbatim}
    def getA(self, freq):
        MfMui = self.MfMui
        MeSigma = self.MeSigma
        C = self.mesh.edgeCurl
        return C.T*MfMui*C + 1j*omega(freq)*MeSigma

    def getRHS(self, freq):
        s_m, s_e = self.getSourceTerm(freq)
        MfMui = self.MfMui
        C = self.mesh.edgeCurl
        return C.T * (MfMui * s_m) -1j * omega(freq) * s_e
\end{verbatim}
\end{multicols}
}
\noindent
and associated methods \texttt{getADeriv} and \texttt{getRHSDeriv} to
construct the derivatives of each with respect to the inversion model. These
function definitions are methods of the \Problem class, where the
\texttt{self} variable refers to the instance of the class, and is standard
Python (cf. Python documentation -
https://docs.python.org/3/tutorial/classes.html). For
\FDEM \texttt{Problem\_e}, \texttt{getRHSDeriv} is zero unless one or both of
the source terms have model dependence. However, if we eliminate $\mathbf{e}$ and solve for $\mathbf{b}$ (\texttt{Problem\_b}), the
right hand side contains the matrix $\MeSigma$, and therefore will, in
general, have a non-zero derivative. To solve this linear system of equations,
\SimPEG interfaces to standard numerical solver packages (e.g. SciPy, Mumps
\citep{scipy, Amestoy2001, Amestoy2006}, using for example pymatsolver
https://github.com/rowanc1/pymatsolver). The components used to perform the forward simulation
are assembled in the
\texttt{fields} method of the \texttt{BaseFDEMProblem} class; the \texttt{fields} method solves the forward simulation for the solution vector $\mathbf{u}$ (from equation
\ref{eq:problem_e}) at each frequency and source considered.

Similarly for the time-domain problem, the semi-discretized E-B formulation is given by
\begin{equation}
    \begin{split}
        \dcurl \mathbf{e} + \frac{d \mathbf{b}}{dt} = \sm \\
        \dcurl^\top \MfMui \mathbf{b} - \MeSigma \mathbf{e} = \se
    \end{split}
    \label{eq:EB_TDEM}
\end{equation}
and the semi-discretized H-J formulation is given by
\begin{equation}
    \begin{split}
        \dcurl^\top \MfRho \mathbf{j} + \frac{d \MeMu\mathbf{h}}{dt} = \sm \\
        \dcurl \mathbf{h} - \mathbf{j} = \se.
    \end{split}
    \label{eq:HJ_TDEM}
\end{equation}

For the time discretization, we use Backward Euler (cf. \cite{Ascher2008}). To
form the \TDEM \texttt{Problem\_b}, we eliminate $\mathbf{e}$ from equation
\ref{eq:EB_TDEM} and apply Backward Euler for the time discretization. A
single timestep takes the form
\begin{equation}
    \underbrace{\left( \mathbf{C} \MeSigma^{-1} \mathbf{C}^\top \MfMui + \frac{1}{\Delta t^{k}} \right)}_{\mathbf{A}^{k+1}_0(\mathbf{m})}
    \underbrace{\vphantom{\left(\frac{-1}{\Delta t^{k}}\right)}
    \mathbf{b}^{k+1}}_{\mathbf{u}^{k+1}}
    + \underbrace{\vphantom{\left(\frac{-1}{\Delta t^{k}}\right)} \frac{-1}{\Delta t^{k}} \mathbf{I}}_{\mathbf{A}_{-1}^{k+1}(\mathbf{m})}
    \underbrace{\vphantom{\left(\frac{-1}{\Delta t^{k}}\right)}
    \mathbf{b}^{k}}_{\mathbf{u}^{k}}
    = \underbrace{\vphantom{\left(\frac{-1}{\Delta t^{k}}\right)}
    \mathbf{C}\MeSigma^{-1} \mathbf{s_e}^{k+1}
    + \mathbf{s_m}^{k+1}}_{\mathbf{q}^{k+1}(\mathbf{s_m, s_e})}
    \label{eq:TDEM_b_single}
\end{equation}
where $\Delta t^k = t^{k+1} - t^{k}$ is the timestep and the superscripts $k$,
$k+1$ indicate the time index. Each \TDEM problem formulation (ie.
\texttt{Problem\_e}, \texttt{Problem\_b}, \texttt{Problem\_h},
\texttt{Problem\_j}) has methods to create the matrices along the block-diagonals,
$\mathbf{A}_0^{k+1}(\mathbf{m})$ and $\mathbf{A}_{-1}^{k+1}(\mathbf{m})$,
as well as a method to construct the right hand side, $\mathbf{q}^{k+1}(\mathbf{s_m, s_e})$,
at each timestep. When inverting for a model in electrical
conductivity using \texttt{Problem\_b}, the sub-diagonal matrices are
independent of $\mathbf{m}$, however, in other formulations, such as \texttt{Problem\_e},
the sub-diagonal matrices do have dependence on electrical conductivity, thus in general, the model
dependence must be considered. Depending on the solver chosen, it can be
advantageous to make the system symmetric; this is accomplished by multiplying
both sides by $\MfMui^\top$. To solve the full time-stepping problem, we
assemble all timesteps in a lower block bidiagonal matrix, with on-diagonal
matrices $\mathbf{A}_0^k(\mathbf{m})$ and sub-diagonal matrices
$\mathbf{A}_{-1}^k(\mathbf{m})$, giving
\begin{equation}
    \underbrace{
        \begin{pmatrix}
            \mathbf{A}^{0}_{0}(\mathbf{m})  &                                 &                                &                                   &                                   &                                  \\
            \mathbf{A}^{1}_{-1}(\mathbf{m}) & \mathbf{A}^{1}_{0}(\mathbf{m})  &                                &                                   &                                   &                                  \\
                                            & \mathbf{A}^{2}_{-1}(\mathbf{m}) & \mathbf{A}^{2}_{0}(\mathbf{m}) &                                   &                                   &                                  \\
                                            &                                 & \ddots                         & \ddots                            &                                   &                                  \\
                                            &                                 &                                & \mathbf{A}^{n-1}_{-1}(\mathbf{m}) & \mathbf{A}^{n-1}_{0}(\mathbf{m})  &                                  \\
                                            &                                 &                                &                                   & \mathbf{A}^{n}_{-1}(\mathbf{m})   & \mathbf{A}^{n}_{0}(\mathbf{m})   \\
        \end{pmatrix}
    }_{\mathbf{A(m)}}
    \underbrace{
        \begin{pmatrix}
            ~\mathbf{u}^{0~~~}     \\
            ~\mathbf{u}^{1~~~}     \\
            ~\mathbf{u}^{2~~~}     \\
            \vdots~                 \\
            ~\mathbf{u}^{n-1}       \\
            ~\mathbf{u}^{n~~~}
        \end{pmatrix}
    }_{\mathbf{u}}
    =
    \underbrace{
        \begin{pmatrix}
            ~\mathbf{q}^{0~~~}     \\
            ~\mathbf{q}^{1~~~}     \\
            ~\mathbf{q}^{2~~~}     \\
            \vdots~                 \\
            ~\mathbf{q}^{n-1}       \\
            ~\mathbf{q}^{n~~~}
        \end{pmatrix}
    }_{\mathbf{q(s_m, s_e)}}
    \label{eq:TDEM_full}
\end{equation}
When solving the forward simulation, the full time-stepping matrix,
$\mathbf{A}(\mathbf{m})$, is not formed, instead the block system is solved
using forward substitution with each block-row being computed when necessary.
The initial condition, $\mathbf{u}^{0}$, depends on the source type and
waveform; it is computed numerically or specified using an analytic
solution. For example, if using a grounded source and a step-off waveform,
$\mathbf{u}^{0}$ is found by solving the direct current resistivity or the
magnetometric resistivity problem, depending on which field we choose to solve for. When a general current waveform is
considered, the initial condition will be $\mathbf{u}^{0}=\mathbf{0}$, and either
$\mathbf{s_m}$ or $\mathbf{s_e}$, depending on type of the source used, will
have non-zero values during the on-time.

Derivatives of the matrices along the block-diagonals of
$\mathbf{A(m)}$ along with derivatives of the right-hand-side
are stitched together in a forward time stepping approach
to compute the contribution of $\frac{d\mathbf{u}}{d\mathbf{m}}$ to
$\mathbf{Jv}$ and in a backwards time stepping approach for the contribution
of $\frac{d\mathbf{u}}{d\mathbf{m}}^{\top}$ to $\mathbf{J}^{\top}\mathbf{v}$.


\subsection{Sources}
\label{sec:Sources}

Sources input EM energy into the system. They can include grounded wires,
loops, dipoles and natural sources. Controlled sources are implemented in the
\FDEM and \TDEM modules of \simpegEM, and natural sources are implemented in
the \NSEM module. For simulations, we require that the sources be discretized
onto the mesh so that a right-hand-side for the Maxwell system can be
constructed (i.e. \texttt{getRHS}). This is addressed by the \texttt{eval}
method of the source which returns both the magnetic and electric sources
($\sm, \se$, shown in Figure \ref{fig:simpegFwd}) on the simulation mesh.

In some cases, a primary-secondary approach can be advantageous for addressing
the forward problem (cf. \cite{Coggon1971, Haber2014a, Heagy2015}). We split up the fields
and fluxes into primary and secondary components ($\mathbf{e} =
\mathbf{e}^\mathcal{P} + \mathbf{e}^\mathcal{S}$, $\mathbf{b} =
\mathbf{b}^\mathcal{P} + \mathbf{b}^\mathcal{S}$ ) and define a ``Primary
Problem'', a simple problem, often with an analytic solution, that is solved
in order to construct a source term for a secondary problem. For instance, a
point magnetic dipole source may be simulated by defining a zero-frequency
primary which satisfies
\begin{equation}
\begin{split}
    \mathbf{e}^\mathcal{P} = 0 \\
           \dcurl^\top \MfMui^\mathcal{P} \mathbf{b}^\mathcal{P} = \mathbf{s_e}^\mathcal{P}.
\end{split}
\label{eq:mag_dipole_primary}
\end{equation}
If we define $\boldsymbol{\mui}^\mathcal{P}$ to be a constant, equation
\ref{eq:mag_dipole_primary} has an analytic solution for
$\mathbf{b}^\mathcal{P}$ that may be expressed in terms of a curl of a vector
potential (cf. \cite{Griffiths2007}). When using a mimetic discretization, by
defining the vector potential and taking a discrete curl, we maintain that the
magnetic flux density is divergence free as the divergence operator is in the
null space of the edge curl operator ($\nabla \cdot \nabla \times \vec{v} = 0$), so
numerically we avoid creating magnetic monopoles (c.f. \cite{Haber2014a}). The
secondary problem is then
\begin{equation}
\begin{split}
\dcurl \mathbf{e}^\mathcal{S} + i \omega \mathbf{b}^\mathcal{S} = - i \omega \mathbf{b}^\mathcal{P} \\
\dcurl^\top \MfMui \mathbf{b}^\mathcal{S} - \MeSigma \mathbf{e}^\mathcal{S} = -\dcurl^\top \left(\MfMui-\left(\MfMui\right)^\mathcal{P}\right)\mathbf{b}^\mathcal{P}
\label{eq:primsecmagdipole}
\end{split}
\end{equation}
The source terms for the secondary problem are $\sm = -i \omega
\mathbf{b}^\mathcal{P},$ and $\se = -\dcurl^\top (\MfMui-\MfMui^\mathcal{P})
\mathbf{b}^\mathcal{P}$. In scenarios where magnetic permeability is homogeneous,
the electric source contribution is zero.

The left hand side is the same discrete Maxwell system as in equation \ref{eq:EB_FDEM};
the  distinction is that we are solving for secondary fields, and a primary
problem was solved (analytically or numerically) in order to construct the
source terms. To obtain the total fields, which we sample with the receivers,
we must add the primary fields back to the solution. To keep track of the
primary fields, they are assigned as properties of the source class.

In most cases, source terms do not have a derivative with respect to the model.
However, in a primary-secondary problem in electrical conductivity the source
term depends on the electrical conductivity and derivatives must be considered
(see Section \ref{sec:steelcasing}). This is similar to inverting for
magnetic permeability using a primary-secondary approach described in
equation~\ref{eq:primsecmagdipole} \citep{Coggon1971, Haber2014a, Heagy2015}. It is also
possible to consider your inversion model to be the location or waveform of the source,
in which case the derivative is also non-zero and source derivatives can be
included in the optimization procedure.


\subsection{Fields}
\label{sec:Fields}

By solving the second-order linear system, as in equation \ref{eq:problem_e},
we obtain a solution vector, $\mathbf{u}$, of one field or flux everywhere in the
domain. In the case of a primary-secondary problem, this solution is a
\emph{secondary} field. To examine all of the fields, we require easy
access to the total fields and total fluxes everywhere in the domain.
This is achieved through the \texttt{Fields} object.

For efficient memory usage, only the solution vector is stored, all other
fields and fluxes are calculated on demand through matrix vector
multiplications. As such, each problem type
(\texttt{e}, \texttt{b}, \texttt{h}, \texttt{j}) has an associated Fields object
with methods to take the solution vector and translate it to the desired field
or flux. For instance, \texttt{Fields\_j} stores the solution vector from
\texttt{Problem\_j} and has methods to compute the total magnetic field in the
simulation domain by first computing the secondary magnetic field
from the solution vector ($\mathbf{u}$; in this example, $\mathbf{u} = \mathbf{j}$) and adding back any
contribution from the source
\begin{equation}
    \mathbf{h} = \frac{1}{i\omega} \MeMu^{-1} \left(-\mathbf{C}^{\top} \MfRho \mathbf{u} + \mathbf{s_m} \right)
\label{eq:h_from_j}
\end{equation}

For their contribution to the sensitivity
(equation~\ref{eq:generalsensitivity}), the fields have methods to compute
derivatives when provided the vectors $\mathbf{v}$ and
$\frac{d\mathbf{u}}{d\mathbf{m}} \mathbf{v}$ (from the \texttt{Physics}). For instance,
for $\mathbf{h}$
\begin{equation}
 \frac{d\mathbf{h}}{d\mathbf{m}} \mathbf{v} =
 \frac{d\mathbf{h}}{d\mathbf{u}} \left(\frac{d\mathbf{u}}{d\mathbf{m}} \mathbf{v} \right)
        + \left(\frac{d\mathbf{h}}{d\se} \frac{d\se}{d\mathbf{m}}
        + \frac{d\mathbf{h}}{d\sm} \frac{d\sm}{d\mathbf{m}}
        + \frac{\partial\mathbf{h}}{\partial\mathbf{m}} \right)\mathbf{v}
\label{eq:dhdm}
\end{equation}

The derivatives for $\mathbf{e}$, $\mathbf{b}$, and $\mathbf{j}$ take the same
form. Conceptually, the product of the full derivative and a vector
$\left(\frac{d\mathbf{f}}{d\mathbf{m}} \mathbf{v}\right)$ can be thought of as a stacked
vector of all of the contributions from all of the fields and fluxes, however,
this is never formed in practice.


\subsection{Receivers}
\label{sec:Receivers}

The measured data consist of specific spatial components of the fields or
fluxes sampled at the receiver locations at a certain time or frequency.
Receivers have the method \texttt{eval} that interpolates the necessary
components of the fields and fluxes to the receiver locations and evaluates
the data required for the problem, such as the frequency domain fields or
natural source impedance data. For the frequency domain problem, real and
imaginary components are treated as separate data so that when inverting, we
are always working with real values. The separation of the data evaluation
from fields in receiver objects allows the derivative computation to be
performed and tested in a modular fashion; this enables rapid development and
implementation of new receiver types.


\subsection{Summary}
\label{sec:Summary}

Having defined the role of each of the elements in the forward simulation
framework outlined in Figure~\ref{fig:simpegFwd}, the necessary machinery to
compute predicted data and sensitivities is at hand for both \FDEM and \TDEM
problems. The modular nature of the framework allows us to make several
abstractions which make the code more transparent and ensure consistency
across implementations. For instance, the definition of the physical
properties and associated inner product matrices is common to all formulations
in both time and frequency domains. Thus, these are defined as properties of a
\BaseEM class which is inherited by both the \TDEM and \FDEM modules. Within
each of the \TDEM and \FDEM modules, common methods for the calculation of the
fields, sensitivities and adjoint are defined and shared across the approaches
that solve for $\mathbf{e}$, $\mathbf{b}$, $\mathbf{h}$, or $\mathbf{j}$ (see
the documentation http://docs.simpeg.xyz).

Testing is conducted using comparisons with analytics,
cross-comparisons between formulations,
order tests on the sensitivity, adjoint tests, examples, tests on the finite volume operators,
projections, interpolations, solvers, etc. Tests are run upon each
update to the repository through the continuous integration service TravisCI
\citep{Travis}. This ensures that we can trust the tools that we use and move
faster in our research into new methods and implementations. This also
supports new developers and researchers in contributing to the code base without
fear of breaking assumptions and ideas laid out by previous development.


\section{Examples}
\label{sec:Examples}

To demonstrate the application and structure of the framework, we explore three examples, one field example and two synthetic examples.
The purpose of the first synthetic example is to show simple time
and frequency domain electromagnetic inversions, and highlight the common
framework. For this, we invert for a 1D layered Earth using a 2D cylindrically
symmetric mesh for the forward simulation. In the second example, we show 1D inversions of field data (RESOLVE and SkyTEM)
collected over the Bookpurnong Irrigation district in Australia.
The final example is a 3D synthetic example that demonstrates a
sensitivity analysis using a parametric model of a block in a layered space
for a reservoir characterization problem where the transmitter is positioned
down-hole in a steel-cased well. We use this example to demonstrate how
mappings, multiple physical properties (both electrical conductivity and
magnetic permeability), and multiple meshes, a cylindrically symmetric and a
3D tensor mesh, can be composed in a primary-secondary approach for performing
the forward simulation and computing the sensitivities. The scripts used to
run these examples are available on http://docs.simpeg.xyz.


\subsection{Cylindrically Symmetric Inversions}
\label{sec:CylindricallySymmetricInversions}

The purpose of this example is to demonstrate the implementation of the
electromagnetic inversion in both time and frequency domains. We have chosen
this example as it is computationally light, can be run on any modern
laptop without installing complex dependencies, and yet it uses most of the
elements and functionality needed to solve a large 3D EM problem. The script used to
run this simulation is available at: https://doi.org/10.6084/m9.figshare.5035175.

We consider two 1D inversions for log-conductivity from an EM survey, one
frequency domain experiment and one time domain experiment. Both surveys use a
vertical magnetic dipole (VMD) source located on the surface. For simplicity,
we consider a single receiver, measuring the vertical magnetic field, located 50m
radially away from the source. The magnetic permeability is taken to be that of free space ($\mu = \mu_0$), and electrical conductivity is assumed to be frequency-independent.

Figure \ref{fig:example1structure} shows the setup used for: (a) the frequency
domain simulation, (b) the time domain simulation, and (c) the common
inversion implementation. In both, a cylindrical mesh is employed for the
forward simulation and a 1D layered earth, described in terms of
log-conductivity. To map the inversion model to electrical
conductivity, a composite mapping is used to inject the 1D subsurface model
into one including air cells (\texttt{InjectActiveCells}), surject the 1D
model onto the 2D simulation mesh (\texttt{SurjectVertical1D}) and take the
exponential to obtain electrical conductivity (\texttt{ExpMap}), as described
in the Model and Physical Properties section (Section \ref{sec:ModelPhysProps}).

{
\begin{figure}[htb!]
    \centering
    \includegraphics[width=\textwidth]{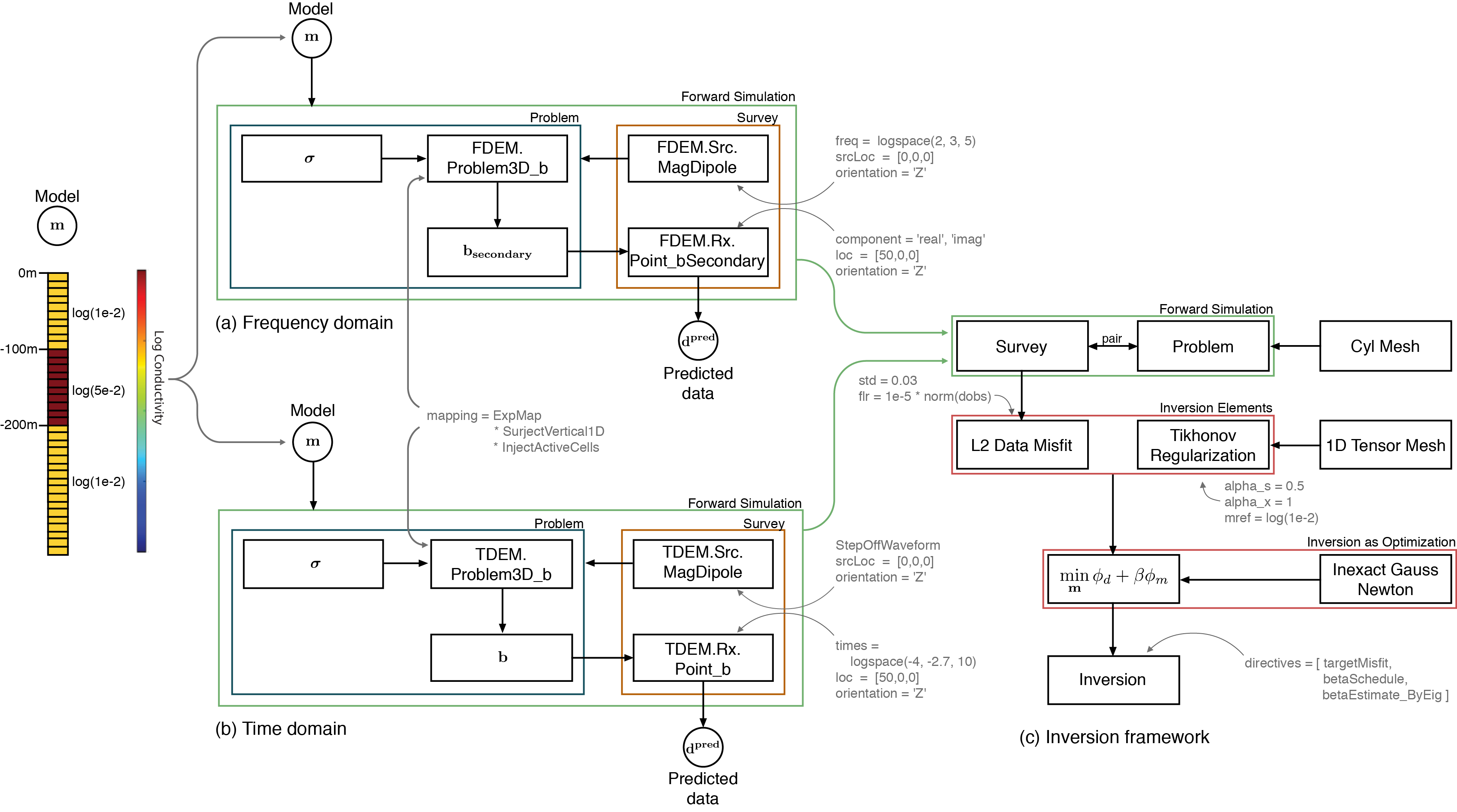}
\caption{Diagram showing the entire setup and organization of (a) the frequency domain simulation; (b) the time domain simulation; and (c) the common inversion framework used for each example. The muted text shows the programmatic inputs to each class instance.}
 \label{fig:example1structure}
\end{figure}
}

The distinction between the frequency and time domain inversions
comes in the setup of the forward simulations. Each employs the appropriate
description of the physics (FDEM or TDEM) in the problem, and the definition
of the survey, consisting of both sources and receivers, must be tailored to
the physics chosen. For the FDEM survey, a vertical harmonic magnetic dipole
located at the origin transmits at five frequencies logarithmically spaced
between 100 Hz and 1000 Hz. The receiver is located at (50 m, 0 m, 0 m)
and measures the secondary magnetic flux (with the primary being the free-space
response of a harmonic magnetic dipole). The observed response is complex-valued,
having both real and imaginary components. We consider these as
separate data, giving a total of ten data points for this example. For the
time domain survey, we again use a vertical magnetic dipole at the origin,
however, we now use a step-off waveform. The observed responses are defined
through time, and thus are all real-valued. For this example, we sample 10
time channels, logarithmically spaced between $10^{-4} ~s$ and $2\times10^{-3}
~ s$ . These time channels were selected to be sensitive to depths similar to
the FDEM simulation.

With the forward simulation parameters defined in both the time and frequency
domain simulations, we can generate synthetic data. The model used consists of
a 100m thick conductive layer (0.05 S/m) whose top boundary is 100 m-below
from the surface, as shown in Figure \ref{fig:example1structure}. The
conductivity of the half-space earth is 0.01 S/m. In both cases, 3\% gaussian
noise is added to the simulated data, and these are treated as the observed
data ($\mathbf{d}^{\text{obs}}$) for the inversion.

For the inversions, we specify the inversion elements: a data misfit and a
regularization. We use an $\ell_2$ data misfit of the form
\begin{equation}
    \phi_d = \frac{1}{2}\|\mathbf{W_d}(\mathbf{d}^{\text{pred}} - \mathbf{d}^{\text{obs}})\|_2^2
    \label{eq:datamisfit}
\end{equation}
where $\mathbf{W_{d_{ii}}} = 1/\epsilon_i$ and we define $\epsilon_i = 3\% |d^{\text{obs}}_i| + \text{floor}$. For both simulations the floor is set to $10^{-5}\| \mathbf{d}^\text{obs}\|$. The regularization is chosen to be a Tikhonov regularization on the 1D model
\begin{equation}
    \phi_m = \frac{1}{2}\left(\alpha_s\|\mathbf{m}- \mathbf{m}_{\text{ref}} \|_2^2 + \alpha_x\| \mathbf{D}_x \mathbf{m}\|_2^2\right)
\end{equation}
where $\mathbf{m}_{\text{ref}}$ is the reference model which is set to be a half-
space of $\log(10^{-2})$. The matrix $\mathbf{D}_x$ is a 1D gradient
operator. For both examples $\alpha_s = 0.5$ and $\alpha_x = 1$. The data
misfit and regularization are combined with a trade-off parameter, $\beta$, in
the statement of the inverse problem. To optimize, we use the second-order
Inexact Gauss Newton scheme. In this inversion we use a beta-cooling
approach, where $\beta$ is reduced by a factor of 4 every 3 Gauss Newton
iterations.

The initial $\beta$ is chosen to relatively weight the influence of the data
misfit and regularization terms. We do this by estimating the largest
eigenvalue of $\mathbf{J}^{\top}\mathbf{J}$ and
$\mathbf{W_m}^{\top}\mathbf{W_m}$ using one iteration of the power method.
We then take their ratio and multiply by a scalar to weight their relative
contributions. For this example, we used a factor of 10. For a stopping
criteria, we use the discrepancy principle, stopping the inversion when
$\phi_d \leq \chi \phi_d^*$, with $\chi = 1$ and $\phi_d^* = 0.5 N_{data}$
(with $\phi_d$ as defined in equation \ref{eq:datamisfit}.)

The FDEM inversion reaches the target misfit after 9 iterations, and the TDEM
inversion reaches the target misfit after 6 iterations.  Figure
\ref{fig:example1Results} shows the recovered models (a), predicted and
observed data for the FDEM inversion (b) and predicted and observed data for
the TDEM inversion (c). In both the FDEM and TDEM inversions, the data are fit
well. The recovered models are smooth, as is expected when employing an $\ell_2$,
Tikhonov regularization. Both the location and amplitude of the
conductive layer is well resolved in the FDEM and TDEM inversions.
The structure of both models are comparable, demonstrating
that the information content in both the FDEM and TDEM data are similar. The
recovered model can be improved by many additional techniques that are not
explored here (e.g. using compact norms in the regularization). The \SimPEG
package provides a number of additional directives and regularization modules
which can be useful for this purpose.

{
\begin{figure}[htb!]
    \centering
    \includegraphics[width=\textwidth]{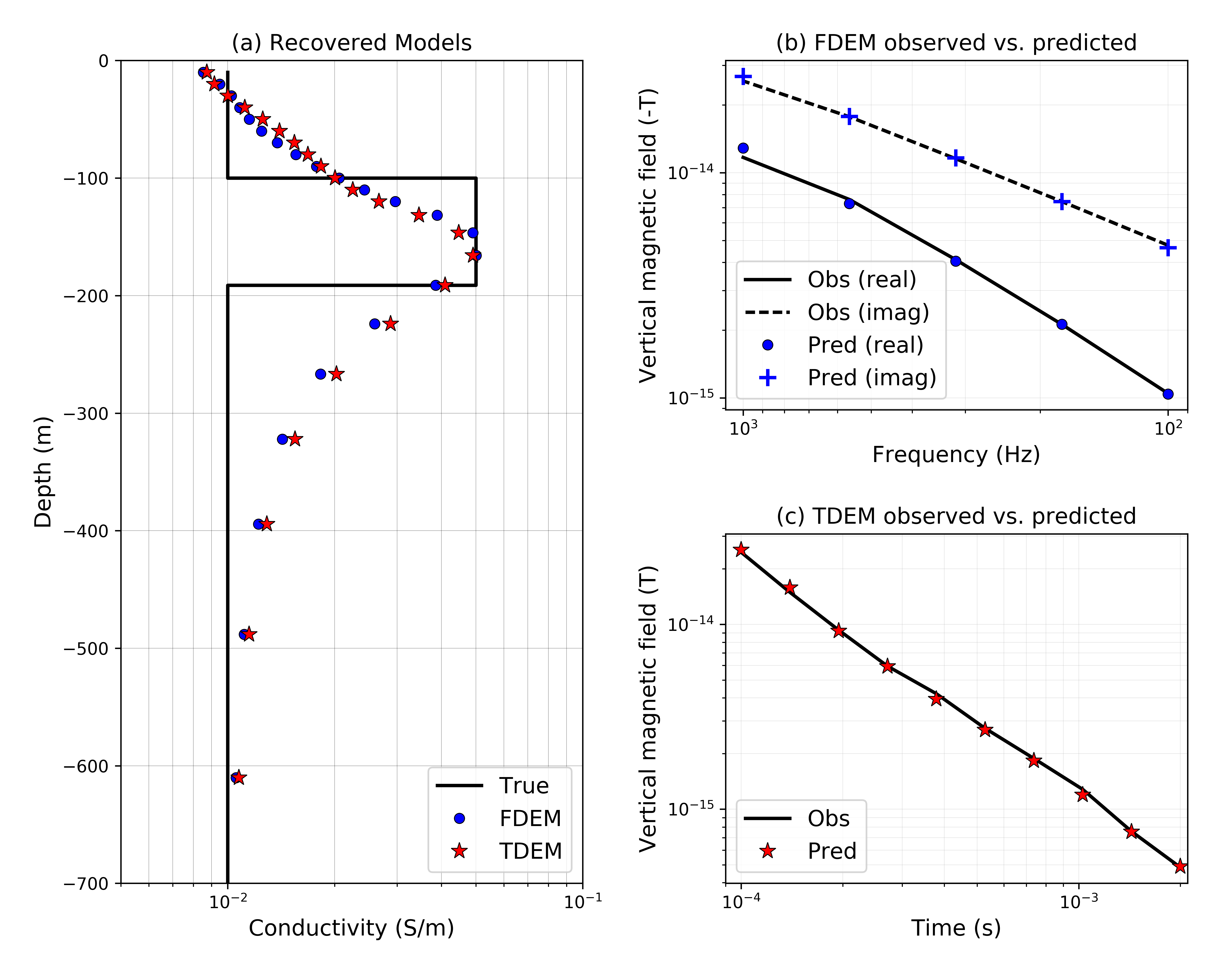}
\caption{(a) True and recovered models for the FDEM and TDEM inversions; predicted and observed data for (b) the FDEM example, and (c) the TDEM example. In (b) the magnetic field data are in the negative z-direction.}
\label{fig:example1Results}
\end{figure}
}


\subsection{Bookpurnong Field Example}
\label{sec:BookpurnongFieldExample}

The purpose of this example is to demonstrate the use of the framework for
inverting field data and provide an inversion that can be compared with other
results in the literature. In particular, we invert frequency and time domain
data collected over the Bookpurnong Irrigation District in Southern Australia.
The Murray River and adjacent floodplain in the Bookpurnong region have become
extensively salinized, resulting in vegetation die-back \citep{Munday2006,
Overton2004}. Multiple electrical and electromagnetic data sets have been
collected with the aim of characterizing the near-surface hydrologic model of
the area \citep{Munday2006}. For a more complete background on the geology and
hydrogeology of the Bookpurnong region, we refer the reader to
\cite{Munday2006}.

Here, we will focus our attention to the RESOLVE frequency-domain data
collected in 2008 and the SkyTEM time-domain data collected in 2006. These data are shown in
Figure \ref{fig:booky_data}. The RESOLVE system consists of 5 pairs of
horizontal coplanar coils, with nominal frequencies of 400 Hz, 1800 Hz, 8200
Hz, 40 000 Hz, and 130 000 Hz as well as a vertical coaxial coil pair of coils which
operates at 3200Hz. For the Bookpurnong survey, the bird was flown at $\sim$50m altitude
\citep{Viezzoli2010}. The SkyTEM time-domain system operates in two
transmitter modes that can be run sequentially. The high moment mode has high
current and operates at a low base frequency (25 Hz and can be lowered to 12.5
Hz), and the low moment operates at a lower current and higher base frequency
(222.5 Hz) \citep{Sorensen2004}. The Bookpurnong SkyTEM survey was flown at an
altitude of $\sim$60m \citep{Viezzoli2010}.

Multiple authors have inverted these data sets; 1D spatially constrained
inversions of the SkyTEM and RESOLVE data were performed by
\citep{Viezzoli2009, Viezzoli2010}. \cite{Yang2017} independently inverted
these data in 1D and provides a discussion at
http://em.geosci.xyz/content/case\_histories/bookpurnong/index.html. The
SkyTEM data (high moment) were inverted in 3D by \citep{Wilson2010}. In the
example that follows, we select a location where both the RESOLVE and SkyTEM
datasets have soundings and invert them in 1D, we then proceed to perform a stitched 1D
inversion of the RESOLVE data. The data have been made available with the
permission of CSIRO and are accessible, along with the script used to run the
inversions at https://doi.org/10.6084/m9.figshare.5107711.

{
\begin{figure}[htb!]
    \centering
    \includegraphics[width=\textwidth]{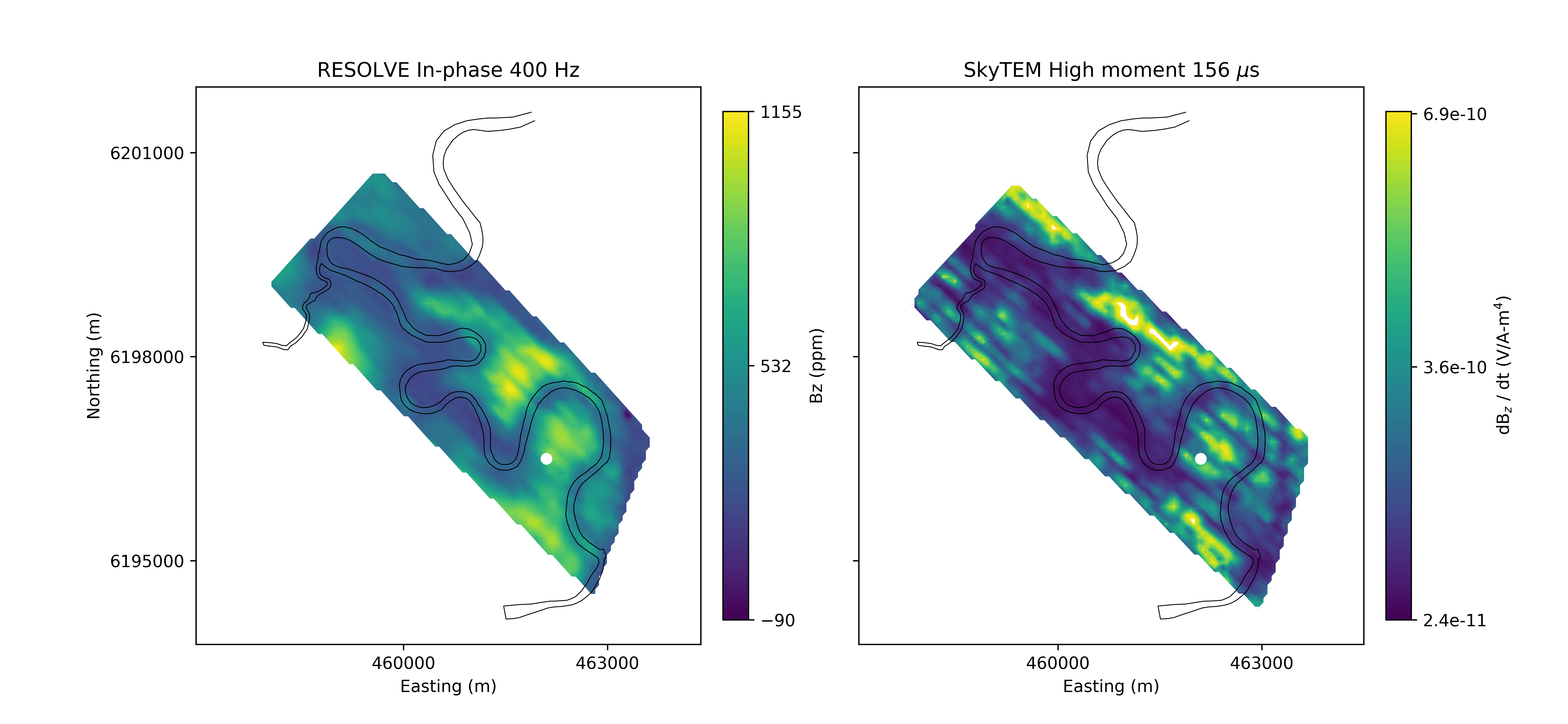}
\caption{
    400 Hz In-phase RESOLVE data at (left) and High Moment SkyTEM data at 156
    $\mu$ s. The white dot at (462100m, 6196500m) on both images is the
    location of the stations chosen to demonstrate the 1D inversions in
    frequency and time.
}
\label{fig:booky_data}
\end{figure}
}

\subsubsection{1D Inversion of RESOLVE and SkyTEM soundings}

We have selected a sounding location (462100m, 6196500m) at which to perform
1D inversions of the RESOLVE and SkyTEM (High Moment) data. The observed data
at this location are shown in Figure \ref{fig:booky1D} (b) and (c). For the RESOLVE inversion, we
consider the horizontal co-planar data collected at 400 Hz, 1800 Hz, 8200 Hz,
40 000 Hz, and 130 000 Hz. For the noise model, we assign 10\% error for the
three lowest frequencies and 15\% error for the two highest; a noise floor of
20ppm is assigned to all data. The inversion mesh uses cells that expand
logarithmically with depth, starting at the surface with a finest cell size of 1m. The forward
simulation is carried out on the cylindrically symmetric mesh, similar to the
previous example. In the inversion, we employ a Tikhonov regularization in
which length scales have been omitted in the regularization function. A fixed
trade-off parameter of $\beta = 2$ is used, $\alpha_z$ is set to be 1, and
$\alpha_s$ is ${10^{-3}}$. A half-space reference model with conductivity 0.1
S/m is used, this also served as the starting model for the inversion. The
inversion reached target misfit after 2 iterations. The resulting model and
data fits are shown in Figure \ref{fig:booky1D}. Very close to the surface, we
recover a resistor, while below that, we recover a conductive unit ($\sim$2
S/m). Examining the data (Figure \ref{fig:booky1D}b), we see that the real
components are larger in magnitude than the imaginary, and that with
increasing frequency, the magnitude of the imaginary component decreases while
the real component increases; such behaviour is consistent with an inductive-
limit response, and we thus expect to recover conductive structures in the
model.

For the time domain inversion, we consider the SkyTEM high moment data. We use
the source waveform shown in the inset plot in Figure \ref{fig:booky1D} (c). For
data, we use 21 time channels from 47 $\mu$s to 4.4 ms; the latest three time
channels (5.6ms, 7ms and 8.8 ms) are not included. For data errors, we assign a
12\% uncertainty and a floor of $2.4\times 10^{-14}$ V/Am$^{4}$. We again use
a Tikhonov regularization, here with $\alpha_z = 1$ and $\alpha_s = 10^{-1}$.
The trade-off parameter is $\beta = 20$. A half-space starting model of 0.1
S/m is again employed. For the reference model, we use the model recovered
from the RESOLVE 1D inversion. As we are using the high-moment data, we do not
expect the SkyTEM data to be as sensitive to the near surface structures as
the RESOLVE data. By using the model recovered in the RESOLVE inversion as the
starting model for the SkyTEM inversion, we can assess agreement between the
two and isolate structures that are introduced by the SkyTEM inversion. The
inversion reached the target misfit after 3 iteration and the results are
shown in Figure \ref{fig:booky1D}. At this location, there is good agreement
in the models recovered from the RESOLVE and SkyTEM data, with both supporting
a near-surface resistor and showing a deeper conductive structure.

{
\begin{figure}[htb!]
    \centering
    \includegraphics[width=\textwidth]{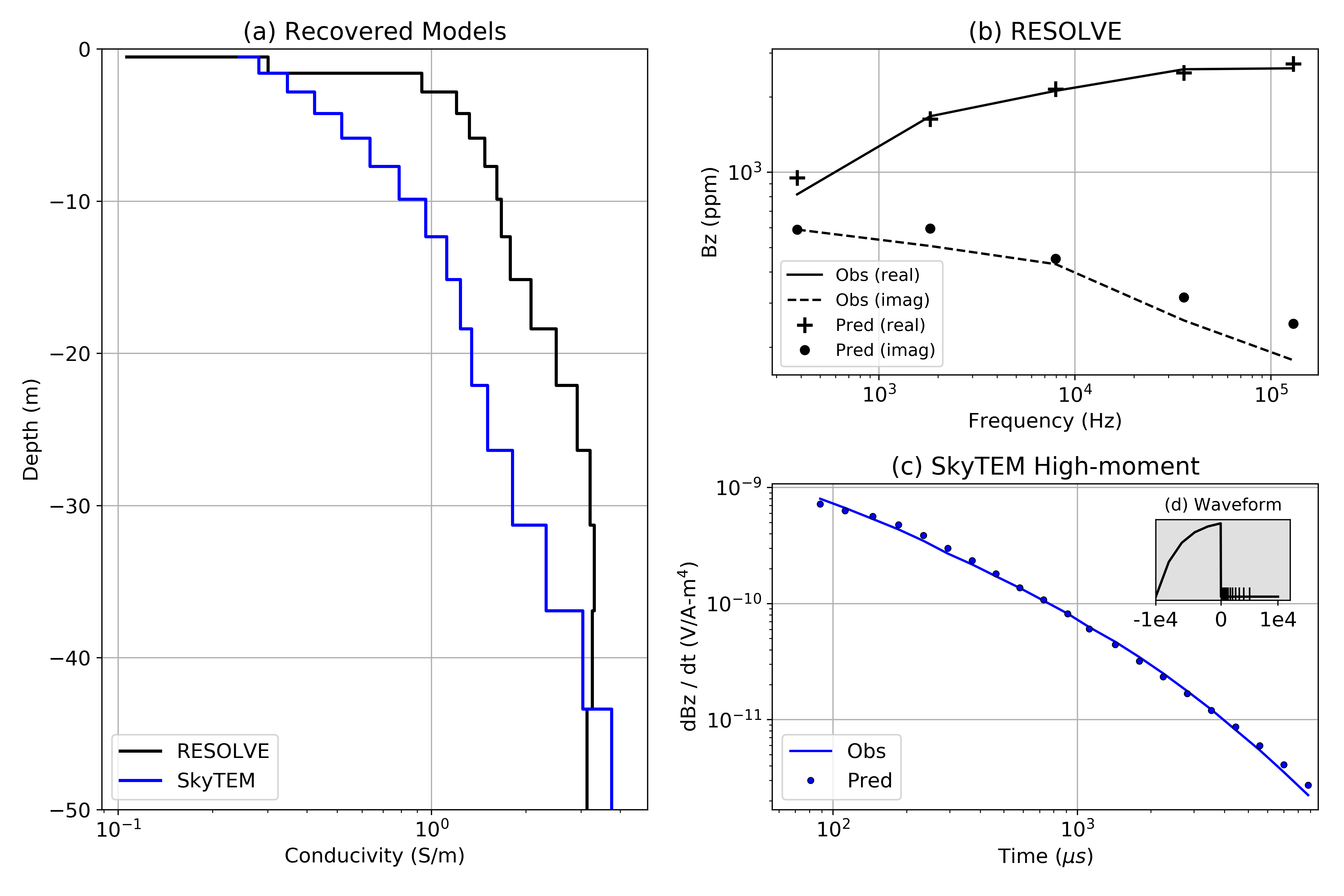}
\caption{
    (a) Models recovered from from the 1D inversion of RESOLVE (back) and SkyTEM
    (blue) data at the location (462100m, 6196500m). (b) Observed (lines) and
    predicted (points) frequency domain data. (c) Observed and predicted time
    domain data. (d) Source waveform used in for the SkyTEM inversion, the x-axis
    is time ($\mu$ s) on a linear scale.
}
\label{fig:booky1D}
\end{figure}
}

\subsubsection{Stitched 1D inversion of RESOLVE data}

Next, we perform a stitched 1D inversion of the RESOLVE data set. With this
example, we aim to demonstrate a practical inversion workflow that will run on
modest computational resources. As such, we have heavily downsampled the data
set, taking 1021 stations of the 40 825 collected. A 1D stitched inversion is
a relatively straight-forward approach for creating a conductivity model -
each sounding is inverted independently and the inversion results are then
assembled to create a 3D model. This can be a valuable quality-control step
prior to adopting more advanced techniques such as including lateral or 3D
regularization across soundings or even performing a 3D inversion. In cases
where the geology is relatively simple, a stitched 1D inversion may be
sufficient. The inversion parameters are the same as those used in the
inversion of the RESOLVE sounding discussed in the previous section. A plan-
view of the recovered model 9.9m below the surface is shown in Figure
\ref{fig:bookyInv}a. A global $\chi$ - factor of 0.74 was reached,
and plots comparing the real component of the observed and
predicted data at 400Hz are shown in Figures \ref{fig:bookyInv} (b) \& (c).

The recovered model (Figure \ref{fig:bookyInv}a), bears similar features to
the models found by \cite{Viezzoli2010} (Figure 4 of \cite{Viezzoli2010}) and by \cite{Yang2017}. In
general, the northwestern portion of the Murray river is more resistive, in
particular near (459 000m, 6 200 000m) and (460 000m, 6 198 000m) while the
southeastern portion of the river is more conductive. Two mechanisms of river
salinization have been discussed in \cite{Munday2006, Viezzoli2010}: the
resistive regions are attributed to a ``losing'' groundwater system, in which
freshwater from the Murray River discharges to adjacent banks, while the
conductive regions are attributed to a ``gaining'' system, in which regional
saline groundwater seeps into the river.

{
\begin{figure}[htb!]
    \centering
    \includegraphics[width=\textwidth]{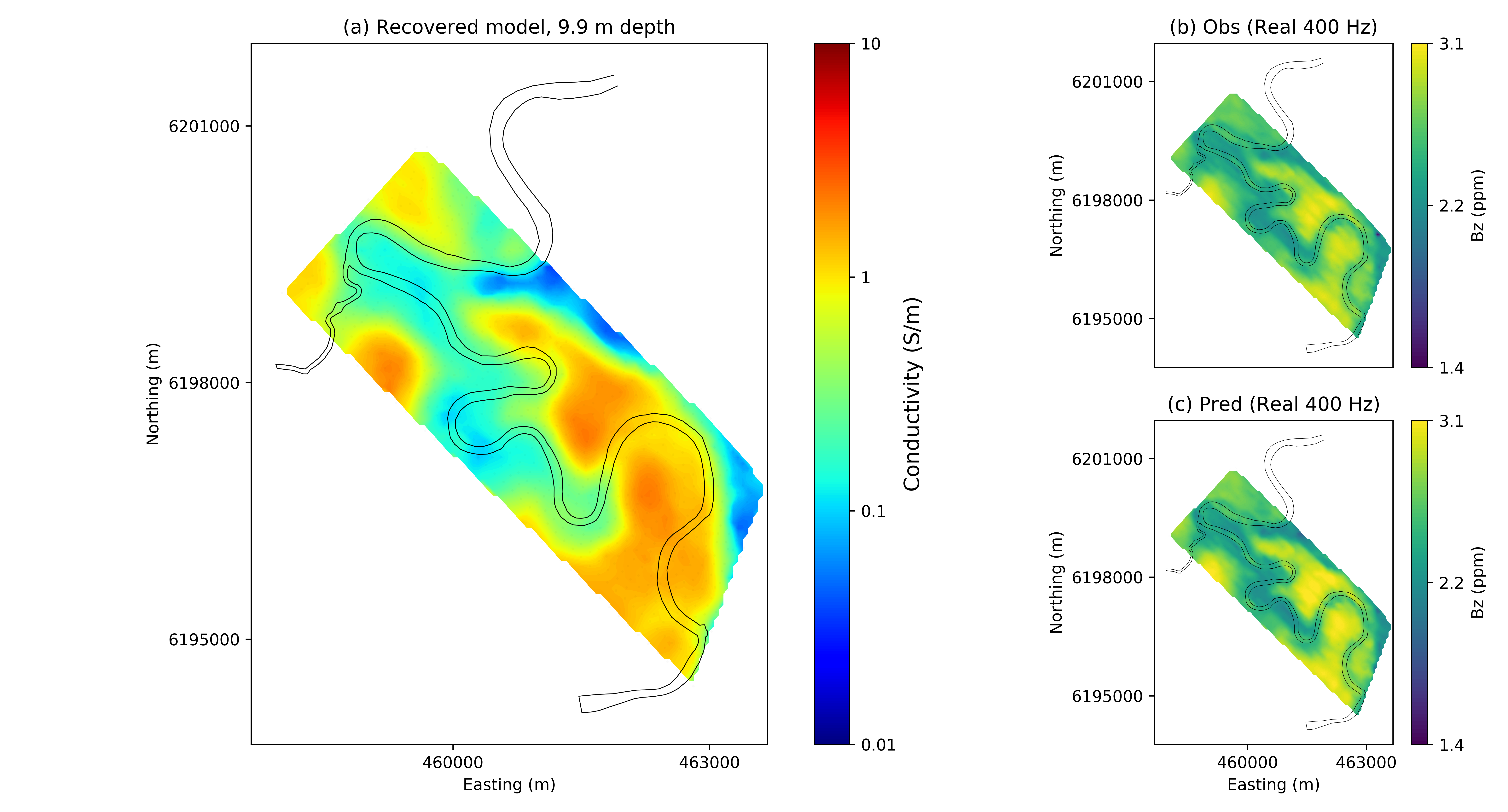}
\caption{
    (a) Conductivity model 9.9m below the surface from a stitched 1D inversion of
    RESOLVE data. (b) Real component of the observed RESOLVE data at 400Hz. (c)
    Real component of the predicted data at 400Hz.
}
\label{fig:bookyInv}
\end{figure}
}


\subsection{Steel-Cased Well: Sensitivity Analysis for a Parametric Model}
\label{sec:steelcasing}

The purpose of this example is to demonstrate the modular implementation of
simpegEM and how it can be used to experiment with simulation and inversion
approaches.  Conducting electromagnetic surveys in settings where steel casing
is present is growing in interest for applications such as monitoring
hydraulic fracturing or enhanced oil recovery \citep{Hoversten2015, Um2015,
Commer2015, Hoversten2014, Marsala2015, cuevas2014eage, Weiss2015,Yang2016a}. Steel is highly
conductive ($\sim 5.5\times10^6 S/m$), has a significant magnetic permeability
($\sim 50\mu_0 - 100 \mu_0$) \citep{wuhabashy1994}. This is a large
contrast to typical geologic settings, with conductivities typically less than
1 S/m and permeabilities similar to that of free space, $\mu_0$. In addition
to the large physical property contrast, the geometry of well casing also
presents a significant computational challenge. Well casing is cylindrical in
shape and only millimeters thick, while the geologic structures we aim to
characterize are on the scale of hundreds of meters to kilometers. Inverting
electromagnetic data from such settings requires that we have the ability to
accurately simulate and compute sensitivities for models with casing and 3D
geologic variations.  One strategy that may be considered is using a primary-
secondary approach, simulating the casing in a simple background and using
these fields to construct a source for the secondary problem which considers
the 3 dimensional structures of interest \citep{Heagy2015}. Here, we
demonstrate how the framework can be employed to implement this approach and
compute the sensitivities. The parametric representation of the model allows
us to investigate the expected data sensitivity to specific features of the model
such as the location, spatial extent and physical properties of a geologic
target. Such an analysis may be used to investigate how well we expect certain
features of the model to be resolved in an inversion and it could be employed as
a survey design tool. In what follows, we outline the general approach and
then discuss a specific implementation. The script used to generate this example is
available at: https://doi.org/10.6084/m9.figshare.5036123.


\subsubsection{Approach}
\label{sec:casingApproach}

In this example we design a survey to resolve a conductive body in
a reservoir layer in the presence of a vertical, steel-cased well as shown in
Figure \ref{fig:parametricCasing}. To calculate the sensitivity of the data
with respect to each model parameter requires that we be able to simulate and
calculate derivatives of each component used to simulate data.

{
\begin{figure}[htb!]
    \centering
    \includegraphics[width=\textwidth]{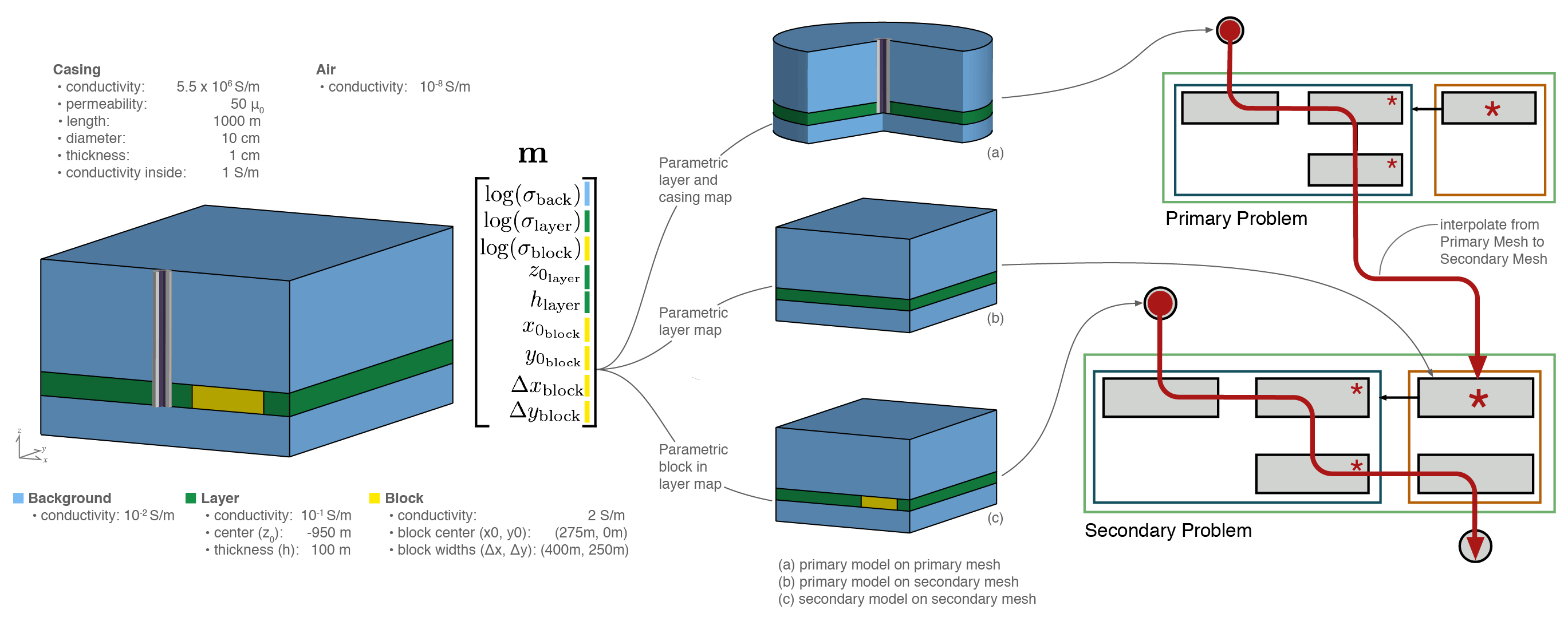}
\caption{Setup of parametric models and calculation of the sensitivity for a primary secondary approach of simulating 3D geology and steel casing.}
\label{fig:parametricCasing}
\end{figure}
}
We use a primary-secondary approach, as described in \cite{Heagy2015}. The
physical properties, fields and fluxes are composed of two
parts, a primary and a secondary part. For example in the E-B formulation, $\sigma
=\sigma^{\mathcal{P}} + \sigma^{\mathcal{S}}$, $\mu =\mu^{\mathcal{P}} +
\mu^{\mathcal{S}}$, $\vec{E} = \vec{E}^{\mathcal{P}} + \vec{E}^{\mathcal{S}}$,
$\vec{B} = \vec{B}^{\mathcal{P}} + \vec{B}^{\mathcal{S}}$. A primary problem,
which includes the cylindrically symmetric part of the model (casing, source, and
layered background) is defined
\begin{equation}
\begin{split}
    \curl \vec{E}^{\mathcal{P}} + i \omega \vec{B^{\mathcal{P}}} = 0 \\
    \curl {\mu^{-1}}^{\mathcal{P}} \vec{B}^{\mathcal{P}} - \sigma^{\mathcal{P}}  \vec{E}^{\mathcal{P}} = \vec{s_e}.
\end{split}
\label{eq:primary}
\end{equation}
This primary problem is solved on a cylindrically symmetric mesh with cells
fine enough to capture the width of the casing and its solution yields the primary fields. The
primary fields are then interpolated to a 3D tensor mesh, suitable for
discretizing 3D reservoir-scale features. The primary fields are used to
construct the source current density for the secondary problem, given by
\begin{equation}
\begin{split}
    \curl \vec{E}^{\mathcal{S}} + i \omega \vec{B}^{\mathcal{S}} = 0 \\
    \curl \mu^{-1}\vec{B}^{\mathcal{S}} - \sigma \vec{E}^{\mathcal{S}} = \vec{q} \\
    \vec{q} =  (\sigma - \sigma^{\mathcal{P}}) \vec{E}^{\mathcal{P}}.
\end{split}
\label{eq:secondary}
\end{equation}
By solving the secondary problem, we then obtain secondary fields and fluxes.
These are sampled by the receivers to create predicted data.

In equation \ref{eq:secondary}, we see that the source term, $\vec{q}$ has
model dependence through $\sigma$, $\sigma^{\mathcal{P}}$ and
$\vec{E}^{\mathcal{P}}$. Typically primary-secondary approaches are used when
the background is assumed to be known, as it is captured in the primary. Here,
however, we do not wish to assume that the background is known; in practice
it may be constrained, but it is not generally well known. The primary solution
is used instead to separate the contributions of the casing and the block
so that we can avoid a potentially crippling assumption. This approach allows
an appropriately tailored mesh to be constructed for each problem. Thus, we
require derivatives not only on the 3D secondary mesh, but also derivatives of
the primary fields (in this case on a cylindrically symmetric mesh). To
implement this type of primary-secondary problem, we construct a
Primary-Secondary source which solves the primary problem to provide the primary
fields. Since all derivatives are implemented for the primary problem, when
computing sensitivities for the secondary problem, the derivatives due to the
primary problem are accounted for in the contributions of the source term to
the derivative. This is conceptually shown in Figure \ref{fig:parametricCasing}.

For this example, we wish to investigate how sensitive the specified survey is
to aspects of the model which we might want to resolve in a field survey, such
as the geometry and location of the anomalous body, as well as the physical
properties of the geologic units. A  voxel-based description of the model does
not promote investigation of these questions, so we will instead apply a
parametric description of the model. The model is parameterized into nine
parameters which we consider to be unknowns
($\log(\sigma_{\text{background}})$, $\log(\sigma_{\text{layer}})$,
$\log(\sigma_{\text{block}})$, $z_{0 _{\text{layer}}}$, $h_{\text{layer}}$,
$x_{0_{\text{block}}}$, $\Delta x_{\text{block}}$, $y_{0_{\text{block}}}$,
$\Delta y_{\text{block}}$). In what follows, we examine the sensitivity of the
data with respect to these model parameters.


\subsubsection{Implementation}
\label{sec:casingImplementation}

The model we use is shown in Figure \ref{fig:parametricCasing}. It  consists of a
1km long vertical steel cased well (diameter: 10 cm, thickness: 1cm) with
conductivity $\sigma= 5.5\times10^6$ S/m, and magnetic permeability $\mu= 50
\mu_0$. The casing is assumed to be filled with fluid having a conductivity of
$1$S/m. The background has a resistivity of $100\Omega \text{m}$, and the 100m
thick reservoir layer has a resistivity of $10\Omega \text{m}$. The target of this
survey is the conductive block ($2$S/m) with dimensions $400 \text{m} \times
250 \text{m} \times 100\text{m}$. The source used consists of two grounded
electrodes, a positive electrode coupled to the casing at a depth of 950m,
and a return electrode 10km from the wellhead on the surface. We consider a
frequency-domain experiment at a transmitting frequency of 0.5Hz and 1A
current. For data, we consider two horizontal components ($x$ and $y$) of the
real part of the electric field measured at the surface.

To accomplish this simulation and sensitivity calculation, we construct 3
mappings, shown conceptually in Figure~\ref{fig:parametricCasing},
in order to obtain: (1) $\sigma^\mathcal{P}$ on the primary (cylindrical) mesh,
(2) $\sigma^\mathcal{P}$ on the secondary mesh (as is needed in equation
\ref{eq:secondary}) and (3) $\sigma$ on the secondary mesh. Differentiability of
the electrical conductivity models with respect to each of the 9 parameters is
achieved by constructing the model using arctangent functions (cf.
\cite{Aghasi2011, McMillan2015}). Each of these parameterizations can be
independently tested for second-order convergence to check the validity of the
computation of the derivatives (cf. \cite{Haber2014a}).

The source term for the secondary fields requires that we simulate the primary
fields. For this, we use the mapping of $\mathbf{m}$ to $\sigma^\mathcal{P}$
on the primary mesh and employ the H-J formulation of Maxwell's equations in
the frequency domain in order to describe a vertically and radially oriented
current density and a rotational magnetic field. In this simulation, we also
consider the permeability of the casing. The source consists of a wire-path
terminating downhole at -950m where it is coupled to the casing. At the
surface, the return electrode is 10km radially away from the well\footnote{Due
to the symmetry employed, the return electrode is a disc. Numerical
experiments over a half-space show that the real, radial electric field from
the cylindrical simulation exhibits the same character as the 3D simulation
but is slightly reduced in magnitude.}. With these parameters defined, we have
sufficient information to solve the primary problem and thereby obtain the
primary electric field everywhere in the simulation domain. The real, primary
current density for this example is shown in Figure
\ref{fig:casingPrimaryFields}.

{
\begin{figure}[htb!]
    \centering
    \includegraphics[width=0.5\textwidth]{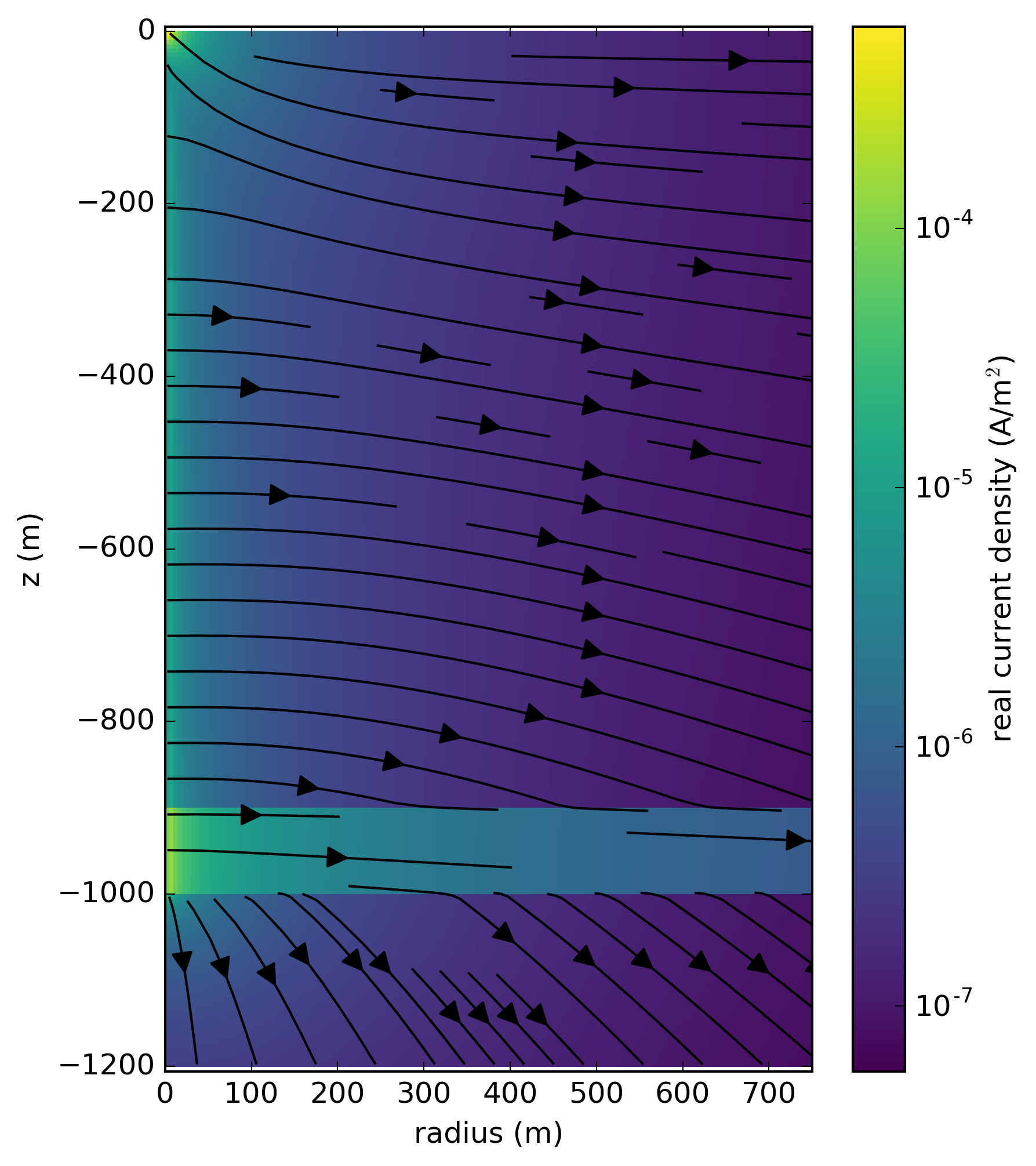}
\caption{Cross sectional slice of primary (casing + background) real current density. The colorbar is logarithmically scaled and shows the amplitude of the real current density. }
\label{fig:casingPrimaryFields}
\end{figure}
}

This primary field is described on the cylindrical mesh, so in order to use it
to construct the source term for the secondary problem, we interpolate
it to the 3D tensor mesh. The remaining pieces necessary for the definition of
the secondary source on the 3D mesh are defining $\sigma$ and $\sigma^\mathcal{P}$; this
is achieved through the mappings defined above. The primary problem and
source, along with the mapping required to define $\sigma^\mathcal{P}$, are used to
define a primary-secondary source, which solves a forward simulation to
compute the secondary source-current, $\mathbf{s_e}$, shown in Figure
\ref{fig:casingSecondary}. Note that the source current density is only
present where there are structures in the secondary model that were not
captured in the primary, in this case, where the conductive block is present.

{
\begin{figure}[htb!]
    \centering
    \includegraphics[width=0.5\textwidth]{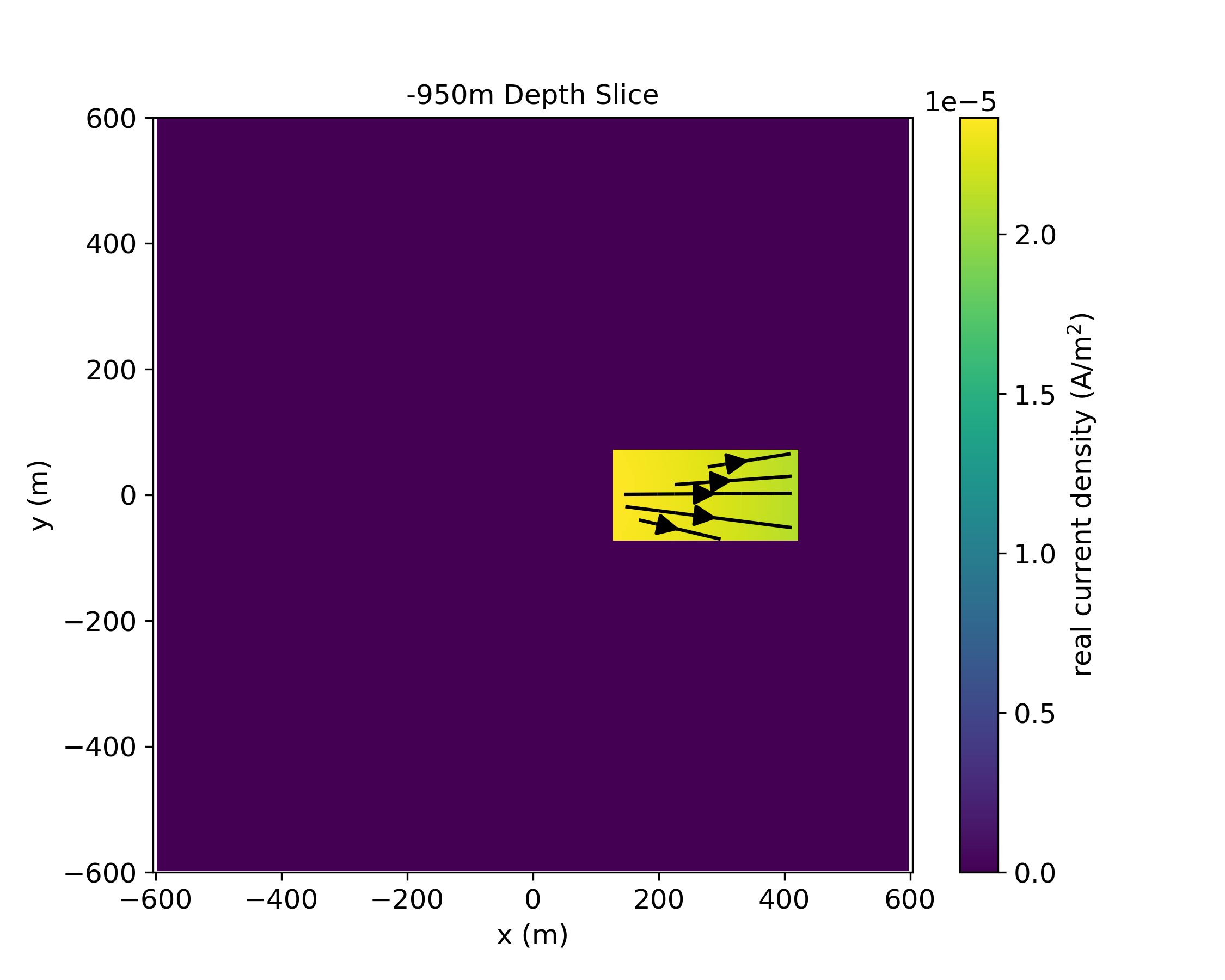}
\caption{Depth slice at z=-950m showing the source current density for the secondary problem.}
\label{fig:casingSecondary}
\end{figure}
}

With the source term for the secondary problem defined, the secondary problem
is then solved resulting in the predicted data at the surface. Here, we focus
our attention to the real x, y components of the electric field, as shown in
Figure \ref{fig:casingData}. The top two panels show the total (casing and
conductive target) x-component (a) and y-component (b)  of the electric field
while the bottom two panels show the secondary (due to the conductive target,
outlined in white) x-component (c) and y-component (d) of the electric field.
As expected, the total electric field is dominated by the source that is
located in the casing. As shown in Figure \ref{fig:casingPrimaryFields} the
majority of the current is exiting into the layer at depth, but current is
still emanating along all depths of the casing. Measured electric fields at
the surface are sensitive to the currents that come from the top part of the
casing and hence the observed fields are strongest closest to the pipe and
they fall off rapidly with distance. The behavior of the secondary electric
field is, to first order, like that expected from a dipole at depth oriented
in the x-direction. It has a broad smooth signature at the surface.

{
\begin{figure}[htb!]
    \centering
    \includegraphics[width=0.77\textwidth]{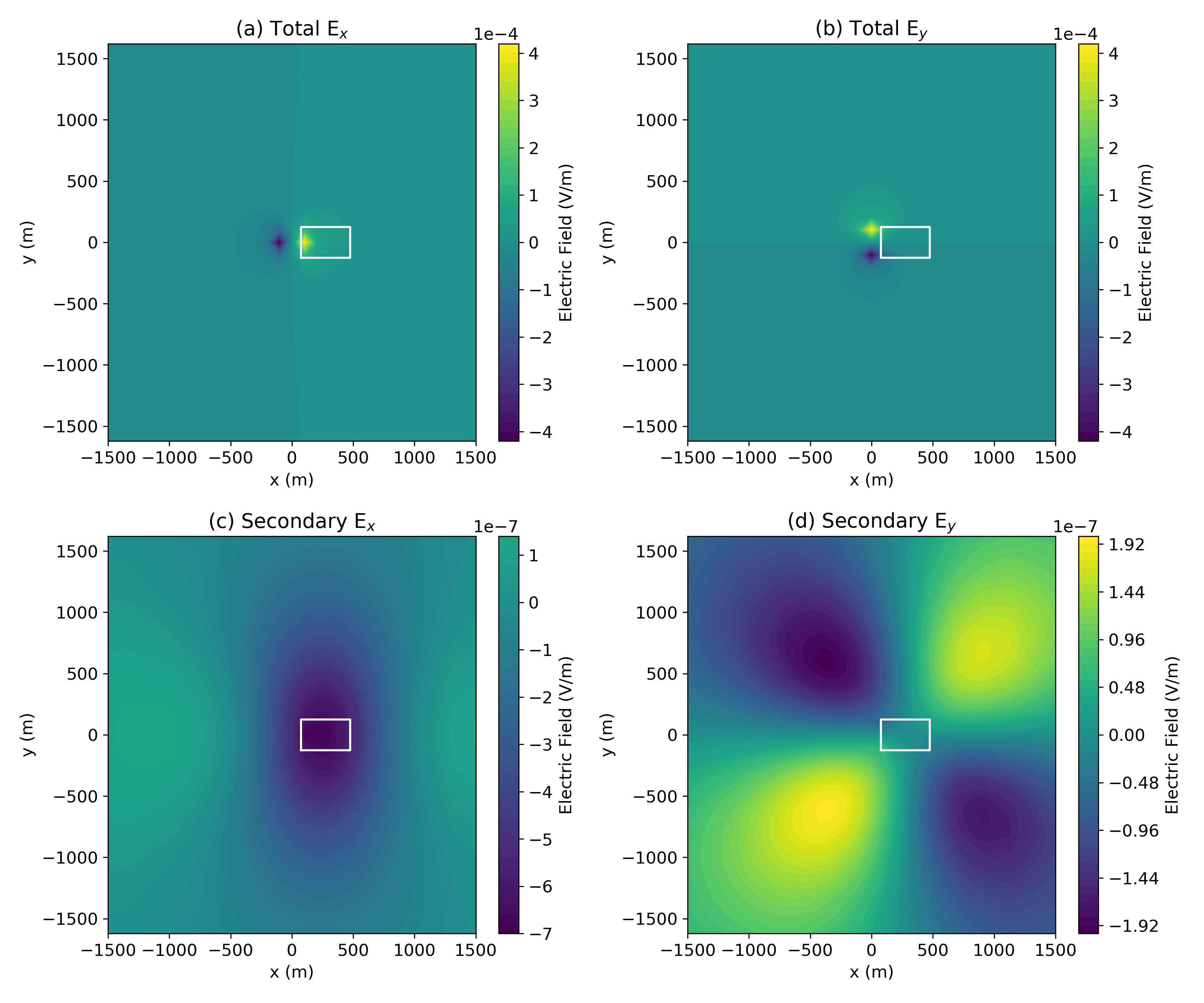}
\caption{Simulated real electric field data as measured at the surface using a
         primary secondary approach for casing and a conductive target (outlined
         in white). The upper panels show the total $E_x$ (a) and $E_y$ (b); the
         lower panels show the secondary (due to the conductive block) $E_x$ (c)
         and $E_y$ (d). Note that the colorbars showing the secondary electric
         fields are not on the same scale. The limits of the colorbars have been set
         so that the zero-crossing is always shown in the same color.}
\label{fig:casingData}
\end{figure}
}

Now that the pieces are in place to perform the forward simulation, we want to
compute the sensitivity. Generally, we do not form the full sensitivity when
performing an inversion as it is a large, dense matrix. Here however, since
the inversion model is composed of only nine parameters, the final sensitivity
matrix is small (nine by number of data). The steps followed to stitch together
and compute the sensitivity are shown in the diagram in Figure
\ref{fig:parametricCasing}. To check the simulation approach for this example,
the sensitivity is tested for second-order convergence (cf.
\cite{Haber2014a}).

Figures \ref{fig:J_sigmas}, \ref{fig:J_layer} and \ref{fig:J_block} shows the
sensitivity of both the real $E_x $(left), and real $E_y$ (right) data with
respect to each of the 9 model parameters. Note that the colorbars are not
identical in each image and the units of the sensitivity are dependent on the
parameter under consideration. In each image, the white outline shows the
horizontal location of the block.

{
\begin{figure}[htb!]
    \centering
    \includegraphics[width=0.77\textwidth]{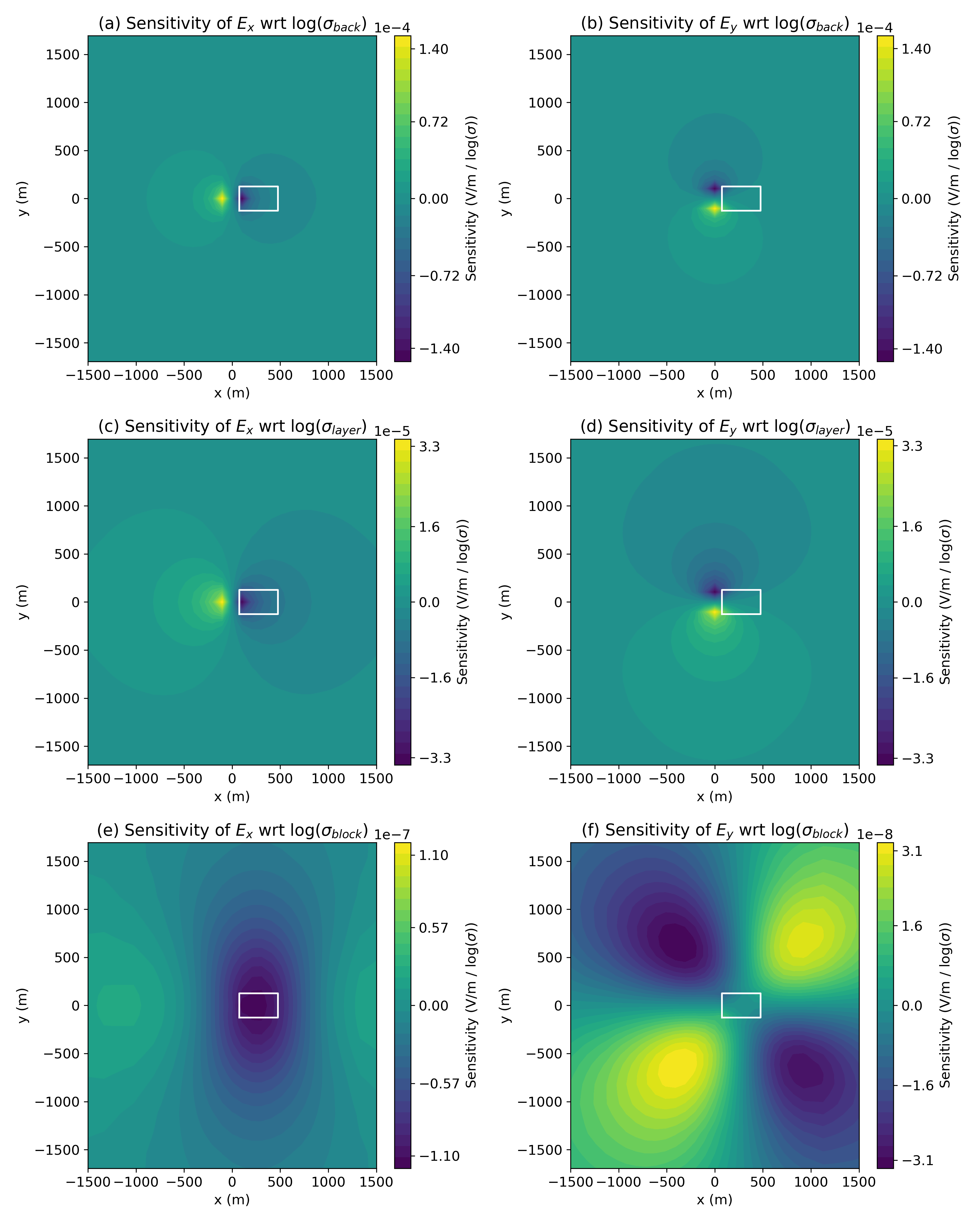}
\caption{Sensitivity of surface real $E_x$ (left) and $E_y$ (right) data with respect to the physical properties, ($(V / m) / (\log(\sigma))$)}
\label{fig:J_sigmas}
\end{figure}
}

{
\begin{figure}[htb!]
    \centering
    \includegraphics[width=0.77\textwidth]{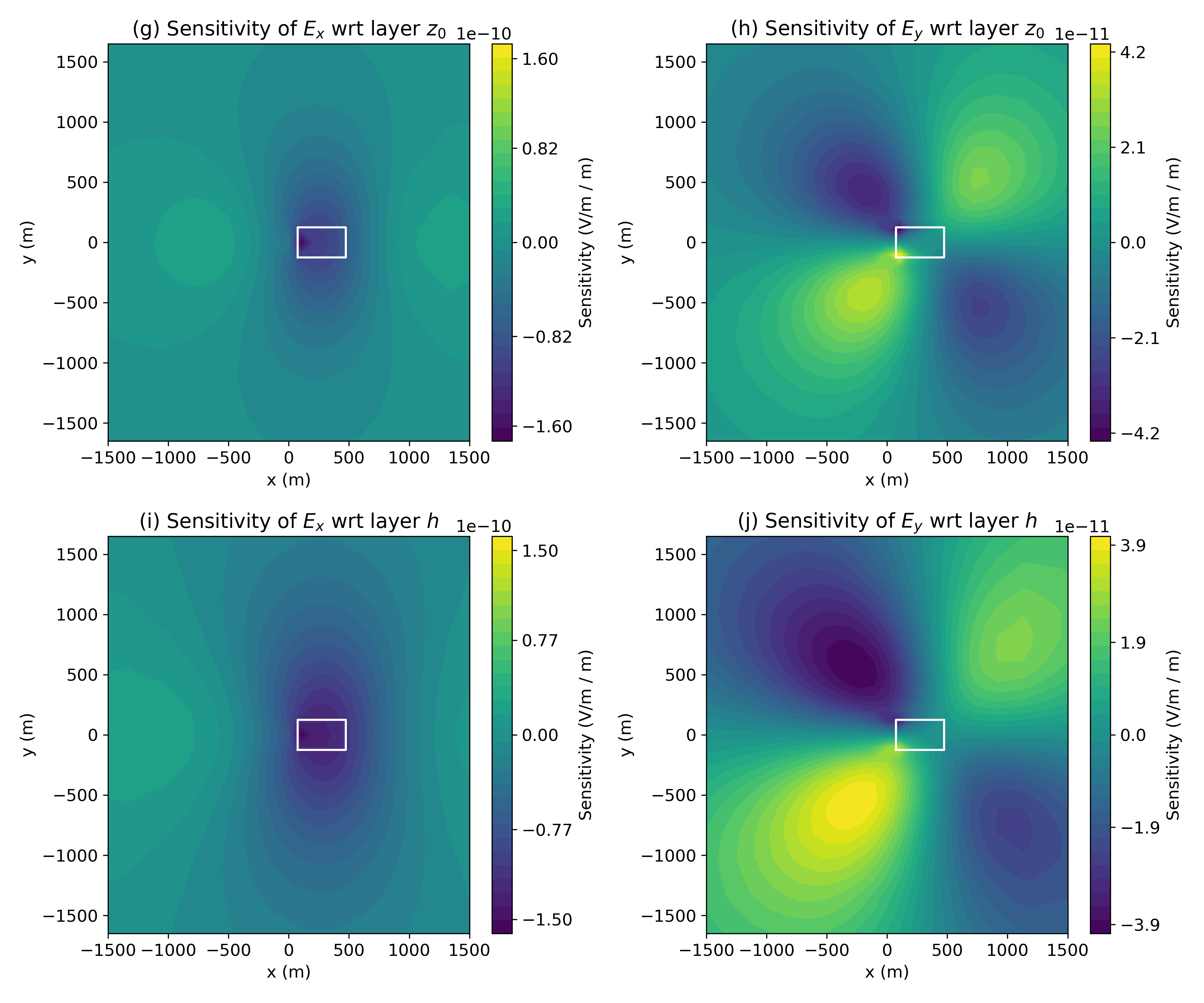}
\caption{Sensitivity of surface real $E_x$ (left) and $E_y$ (right) data with respect to the layer geometry, ($(V /m) / m$)}
\label{fig:J_layer}
\end{figure}
}

{
\begin{figure}[htb!]
    \centering
    \includegraphics[width=0.77\textwidth]{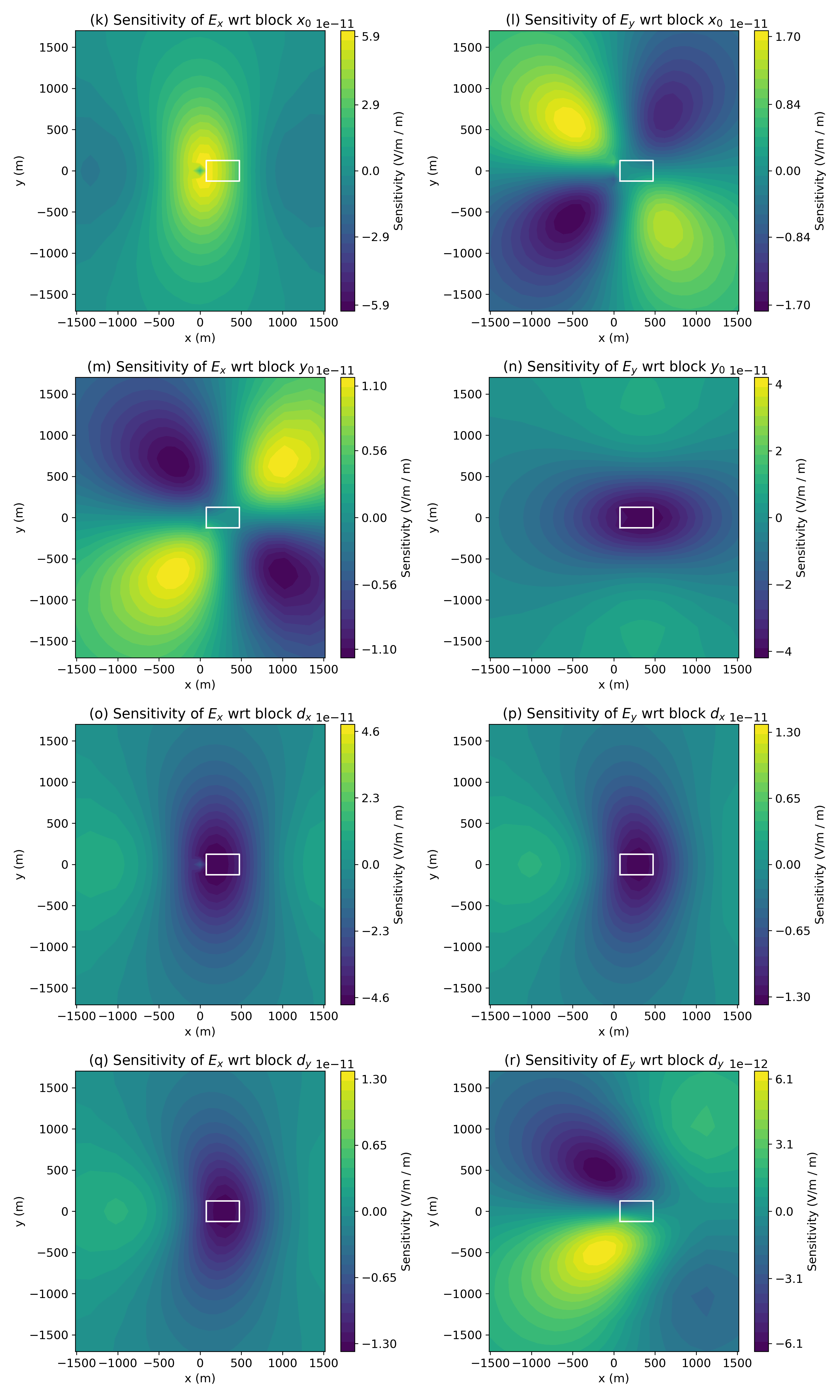}
\caption{Sensitivity of surface real $E_x$ (left) and $E_y$ (right) data with respect to the block geometry, ($(V /m) / m$)}
\label{fig:J_block}
\end{figure}
}

In Figure \ref{fig:J_sigmas}, we focus on the physical properties of the
background layer and block, all parametrized in terms of $\log(\sigma)$.
Clearly, the conductivity of the background has the largest influence on the
data, in particular near the well (at the origin), followed by the
conductivity of the layer, where the injection electrode is situated. There
are 4 orders of magnitude difference between the maximum sensitivity of the data with
respect to the conductivity of the block and that of the background. This
indicates that in order to resolve such an anomalous body, the background
must be well-constrained. When looking at Figure \ref{fig:J_sigmas} (f), we
see that the areas of largest sensitivity of the $E_y$ data with respect to the
physical properties of the block are spatially distant from the body and the
well. This  indicates that if one is designing a survey, it may be advantageous to
collect data in these regions as these are also regions where the influence of the properties
of the background are less dominant.

In Figure \ref{fig:J_layer}, we focus on the depth and thickness of the layer.
Note that the depth and thickness of the block are constrained to be the same
as the layer, so the character of the sensitivity is influenced by the
presence of the block. Here, the units of the sensitivity are $(V/m) / m$.
Similarly, Figure \ref{fig:J_block} shows the sensitivity with respect to the
geometric properties of the block.

To compare between the physical properties and geometry of the model, the
scales of interest must be taken into consideration. In Table
\ref{table:casingSensitivities}, we show the maximum amplitude of the
sensitivity with respect to each individual model parameter. From this, we
approximate the sensitivity as linear about the true model and compute the
perturbation required to cause a change of $10^{-9}~V/m$ in the data
($\Delta \mathbf{m}_i = 10^{-9} / \max|\mathbf{J}_i|$). For ease
of comparison, the perturbations in the log-conductivity of the background,
layer, and block were converted to linear conductivity by
\begin{equation}
    \Delta \sigma_{\text{unit}} = \frac{\exp[\log(\sigma)_\text{unit} + \Delta\log(\sigma)_\text{unit}] -
                                        \exp[\log(\sigma)_\text{unit} - \Delta\log(\sigma)_\text{unit}]}{2}.
\label{eq:log2linear}
\end{equation}

In table \ref{table:casingSensitivities}, we see that to cause a perturbation
in the $E_x$ data by $\sim10^{-9}~V/m$, requires a 0.007\% change in the
conductivity of the background, while the conductivity of the block would need
to change by 0.8\% to have a comparable impact in the $E_x$ data.
In comparing between physical properties and geometric features of the model,
we see that a change in the conductivity of
the block by 0.8\% has a similar impact in the $E_x$ data
as moving $x_0$ of the block by $\sim16~m$. For a change in $y_0$ of the block
to have a comparable impact in the $E_x$ data would require that it be perturbed by
$\sim85~m$. However, the $E_y$ data are more sensitive to $y_0$; a perturbation of $\sim24~m$
, about 1/3 of that required in the $E_x$ data, would result in a $\sim10^{-9}~V/m$ change in the
measured responses.

\begin{table}[htb!]
\centering
  \begin{tabular}{| l | r || r | r  r || r | r  r |}
    \hline
    parameter                     & Units of                    & max $|\mathbf{J}_i|$  & \multicolumn{2}{|r|}{perturbation required to}         & max $|\mathbf{J}_i|$ & \multicolumn{2}{|r|}{perturbation required to}         \\
    $\mathbf{m}_i$                & Sensitivity, $\mathbf{J}_i$ & wrt $E_x$             & \multicolumn{2}{|r|}{cause $\pm 10^{-9} V/m$ in $E_x$} & wrt $E_y$            & \multicolumn{2}{|r|}{cause $\pm 10^{-9} V/m$ in $E_y$} \\
    \hline
    $\log(\sigma_{\text{back}})$  & $(V/m) / \log(\sigma) $     & 1.5e-04               &  6.6e-08 $S/m$  & (6.6e-04\%)                         & 1.5e-04              & 6.6e-08 $S/m$                & (6.6e-04\%)              \\ \hline
    $\log(\sigma_{\text{layer}})$ & $(V/m) / \log(\sigma) $     & 3.5e-05               &  2.9e-06 $S/m$  & (2.9e-03\%)                         & 3.4e-05              & 2.9e-06 $S/m$                & (2.9e-03\%)              \\ \hline
    $\log(\sigma_{\text{block}})$ & $(V/m) / \log(\sigma) $     & 1.2e-07               &  1.7e-02 $S/m$  & (8.4e-01\%)                         & 3.3e-08              & 6.1e-02 $S/m$                & (3.1e+00\%)              \\ \hline
    ${z_0}_{\text{layer}}$        & $(V/m) / m $                & 1.7e-10               &  5.8e+00 $m$    &                                     & 4.4e-11              & 2.3e+01 $m$                  &                          \\ \hline
    ${h}_{\text{layer}}$          & $(V/m) / m $                & 1.6e-10               &  6.2e+00 $m$    &                                     & 4.1e-11              & 2.4e+01 $m$                  &                          \\ \hline
    ${x_0}_{\text{block}}$        & $(V/m) / m $                & 6.2e-11               &  1.6e+01 $m$    &                                     & 1.8e-11              & 5.6e+01 $m$                  &                          \\ \hline
    ${y_0}_{\text{block}}$        & $(V/m) / m $                & 1.2e-11               &  8.5e+01 $m$    &                                     & 4.2e-11              & 2.4e+01 $m$                  &                          \\ \hline
    ${\Delta x}_{\text{block}}$   & $(V/m) / m $                & 4.8e-11               &  2.1e+01 $m$    &                                     & 1.5e-11              & 6.6e+01 $m$                  &                          \\ \hline
    ${\Delta y}_{\text{block}}$   & $(V/m) / m $                & 1.4e-11               &  7.3e+01 $m$    &                                     & 6.5e-12              & 1.5e+02 $m$                  &                          \\ \hline
   \end{tabular}
   \caption{Comparison of the maximum amplitude of the sensitivity with respect to each model parameter,
            and the approximate perturbation in that parameter required to produce a $10^{-9}~ V/m$ change
            in the measured data. The conversion from a perturbation in log-conductivity to conductivity
            is given by equation \ref{eq:log2linear}. The perturbation in conductivity is also provided in terms of
            a percentage of the true model conductivity.}
   \label{table:casingSensitivities}
 \end{table}

Examining the nature of the sensitivity with respect to parameters
describing the target of interest provides insight both into how one might
design a survey sensitive to the target, and how well we may be able to
resolve various geometric features or physical properties in the model.
For the example shown here, we see that it may be advantageous to collect data away from the
well and hundreds of meters offset from the block. These are regions where both the
$E_x$ and $E_y$ data have high sensitivity to features of the target
and are distant from the steel-cased well, where we have the highest sensitivity to the background.
Thus, data collected in
these regions may improve our ability to resolve the target of interest.
The parametric definition of the model provides a mechanism for examining how well we might expect
to resolve various aspects of the target, such as its spatial extent.
There are clearly further questions that may be investigated here, including
exploring survey parameters such as the impact of varying the frequency on our
ability to resolve the block, or performing the same analysis for a time-domain survey.
A modular framework, with accessible derivatives, is an asset for exploring
these types of questions.




\section{Conclusion}
\label{sec:Conclusion}

The framework we have laid out has rigorously separated out various
contributions to the electromagnetic equations in both time and frequency domain.
We have organized these ideas into an object oriented hierarchy
that is consistent across formulations and attends to
implementation details and derivatives in a modular way.
The organization of the framework and its associated numerical implementation are designed to reflect the math.
The goal is to create composable pieces such that electromagnetic
geophysical inversions and forward simulations can be explored and experimented
with by researchers in a combinatorial, testable manner.

We strive to follow best practices in terms of software development including
version control, documentation unit testing, and continuous integration. This work
and the \SimPEG project are open-source and licensed under the permissive MIT
license. We believe these practices promote transparency and reproducibility and we hope
that these promote the utility of this work to the wider geophysics community.

{
\section*{Acknowledgments}

\label{sec:acknowledgments}
The authors thank CSIRO for making the Bookpurnong data available, Dikun Yang for conversations on the inversion of those data, and the three anonymous reviewers whose comments improved the quality of the paper.
We also thank the growing community of \SimPEG developers who have contributed to discussions and improvements in the \SimPEG code-base.

The funding for this work is provided through the Vanier Canada Graduate
Scholarships Program.

}

\section*{References}
\bibliographystyle{elsarticle-harv}
\bibliography{lindseyrefs,seogirefs,gudnirefs}
,
\end{document}